%% file: main_arxiv_v1.tex
\documentclass[twocolumn,amsmath,amssymb,floatfix,nature,shownopacs,footinbib,superscriptaddress]{revtex4-1}
\usepackage{etoolbox}
\usepackage{multirow}
\usepackage{enumitem}
\usepackage{subcaption}
\newcounter{tablerow}
\usepackage{tikz}
\newcommand{\mat}[1]{\mathbf{#1}}

\AtBeginEnvironment{tabular}{\setcounter{tablerow}{0}}

\usepackage[utf8]{inputenc}
\usepackage[T1]{fontenc}
\usepackage{amsmath,amssymb,natbib,bm,graphicx,color,hyperref}

\definecolor{colPolymod}{HTML}{DC143C}
\definecolor{colMistry}{HTML}{1D70A2}
\definecolor{colPrem}{HTML}{FFC857}
\definecolor{colLucchini}{HTML}{00897B}

\appto\UrlBreaks{\do\-}

\begin{document}

\title{Synthetic populations for epidemic, demographic and mobility modeling}

\author{Davide Torre}
\affiliation{ISI Foundation, Turin, Italy}

\author{Miguel Artigues}
\affiliation{Instituto de F\'{\i}sica Interdisciplinar y Sistemas Complejos IFISC (CSIC-UIB), E07122 Palma de Mallorca, Spain}

\author{Benjamin Faucher}
\affiliation{Sorbonne Universit\'{e}, INSERM, Sorbonne Public Health Institute, F75012, Paris, France}

\author{Roberto Alcaraz}
\affiliation{Instituto Mediterr\'aneo de Estudios Avanzados IMEDEA (CSIC-UIB), E07190 Esporles, Spain}

\author{Jacopo D'Ignazi}
\affiliation{Universitat Pompeu Fabra, Barcelona, Spain}

\author{Antonia Tugores}
\affiliation{Instituto de F\'{\i}sica Interdisciplinar y Sistemas Complejos IFISC (CSIC-UIB), E07122 Palma de Mallorca, Spain}

\author{Mattia Mazzoli}
\affiliation{ISI Foundation, Turin, Italy}

\author{Daniela Paolotti}
\affiliation{ISI Foundation, Turin, Italy}

\author{Jos\'e J. Ramasco}
\affiliation{Instituto de F\'{\i}sica Interdisciplinar y Sistemas Complejos IFISC (CSIC-UIB), E07122 Palma de Mallorca, Spain}

\author{Eugenio Valdano}
\affiliation{Sorbonne Universit\'{e}, INSERM, Sorbonne Public Health Institute, F75012, Paris, France}

\maketitle

\section*{Abstract}
Synthetic populations are essential for individual-based simulations, such as agent-based models, enabling the evaluation of policy interventions for epidemic control, demographic studies and mobility management, to mention a few. Existing synthetic populations are often designed for specific settings and rarely combine nationwide coverage with realistic and explicit spatial detail. This limits their ability to represent realistic spatial relationships and interplay between households, workplaces, leisure activities, and mobility patterns, reducing models utility for targeted decision-making. 

Here, we present a harmonized dataset of synthetic populations taking as case studies Spain, France, and Italy. Despite heterogeneous census data sources and formats, we release a unified data structure ensuring consistency and interoperability. The synthetic population generation pipeline reproduces household composition, population density, age-sex structure, and occupation distribution at census-tract level, enriched with municipality-level commuting flows and economic activity patterns. Country-specific pre-processing, statistical calibration, and data integration produce coherent demographic and spatial characteristics suitable for large-scale, country-agnostic simulations. 

As the pipeline relies on publicly available census data, it is readily extensible to other countries, supporting high-resolution modeling of infectious diseases, hazard exposure, demographic analysis, and urban mobility.

\section*{Background and summary}

Over the past two decades, the ability to collect, process, and analyze massive datasets on human behavior has profoundly transformed the way researchers investigate complex human systems. Coupled with increasing computational power, this amount of data has opened the door to highly detailed modeling approaches that account for realistic population heterogeneities across spatial and socio-demographic dimensions. Among these approaches, agent-based models (ABMs) and micro-simulations have become key tools to capture the diversity of human behavior and the structured interactions that shape social \cite{caulfield_curley_open-source_2025}, economic \cite{gelb_synthetic_2024}, or epidemiological dynamics \cite{zhu_generating_2024,son_individual-based_2020}.
Synthetic populations are the foundation of ABMs~\cite{beckman1996creating}. At their core, they are databases of agents and spatial entities (\emph{e.g.} buildings and point of interest), each described by a set of attributes deemed relevant to describe the phenomenon under investigation. A well-constructed synthetic population mirrors the structure and statistics of real-world populations, while guaranteeing anonymization and privacy protection of real data subjects. The selection of variables to be included depends both on the research objectives and the granularity of available data. For example, studies on cardiovascular risk \cite{krauland_development_2020,knight_developing_2017} might require attributes such as smoking status, body mass index, or physical activity levels, while research on epidemic spread \cite{zhu_generating_2024,chae_evaluation_2023,son_individual-based_2020,moghadas_simulated_2021,chao_flute_2010,ajelli_comparing_2010,dyson_modelling_2022} might focus on household composition, workplace size, and commuting flows.

Most synthetic populations are built to reflect socio-demographic characteristics, including age, sex, household membership, and sometimes income or education. These features are often available from census data or demographic surveys, making them relatively straightforward to reconstruct. Several projects have developed large-scale synthetic populations at continental \cite{rineer_national_2025} or even global levels \cite{ton_global_2024}, typically using national or subnational aggregates as their baseline. 
Beyond demography, synthetic populations have been enriched with health-related or behavioral data in order to model disease burden \cite{knight_developing_2017,krauland_development_2020} or evaluate public health policies \cite{rineer_national_2025,tatara_agent_based_2024,stansfield_estimating_2023,pearson_stuttard_estimating_2018}.
Synthetic cohorts have been created to project the incidence of chronic diseases such as diabetes or cardiovascular disease \cite{knight_developing_2017,krauland_development_2020}, to evaluate the cost-effectiveness of screening programs \cite{dyson_modelling_2022}, or to assess differential exposure to environmental risk factors \cite{levy_community_wide_2014}.
In the field of urban planning, synthetic populations have been employed to explore transport demand \cite{pagani_remote-s_2022}, housing needs \cite{somanath_activity-based_2024}, and socioeconomic inequalities \cite{gelb_synthetic_2024} at the city level.
Their flexibility makes them a powerful tool for what-if analyses, where researchers and policymakers can test the potential impacts of interventions \textit{in silico} before real implementation.
A particularly important application lies in epidemiology: synthetic populations underpin a wide range of ABMs designed to simulate the transmission of infectious diseases by explicitly modeling the contact networks and mobility patterns of individuals, providing a more realistic model of epidemic spread than classical compartmental models based on homogeneous mixing. They allow researchers to assess the effectiveness of targeted interventions such as school closures \cite{dyson_modelling_2022}, workplace distancing \cite{chae_evaluation_2023}, or vaccination campaigns \cite{moghadas_simulated_2021}. During the COVID-19 pandemic, ABMs played a central role in decision-making by governments and public health agencies \cite{faucher_agent-based_2022,moreno_lopez_anatomy_2021}.

One of the central dimensions of synthetic population construction is its degree of geographical explicitness. At one end of the spectrum, individuals may simply be distributed into households, schools, and workplaces that are associated with a broad administrative unit such as a province or country \cite{ton_global_2024,chae_evaluation_2023,chao_flute_2010,guarino_inferring_2021}. In such cases, the spatial structure of communities is abstract, and assumptions of homogeneous mixing are often made at the city or regional level. While sufficient for certain applications, this coarse representation can lead to unrealistic estimates of contact patterns and mobility flows, especially in heterogeneous urban environments.
More sophisticated efforts have added increased realism by matching the size distributions of households, schools, and workplaces to empirical data, or by allocating them within synthetic built environments derived from road networks \cite{jiang_method_2022} and population density maps \cite{rineer_national_2025}. In some regional studies, real building footprints have been used to place households and institutions \cite{somanath_activity-based_2024}, thereby creating populations that are not only demographically accurate but also spatially consistent with the urban fabric. However, very few projects have scaled this approach up to the national level while maintaining fine-grained geographical resolution and explicitness.

In this work, we present the construction of a large-scale, geographically explicit synthetic population for three European countries: Spain, France and Italy. 
Our approach integrates multiple datasets to create populations that are simultaneously demographically consistent, spatially explicit, and behaviorally realistic. Each synthetic individual is assigned a set of socio-demographic attributes (age, sex, occupation and household membership) and an explicit geographical position on the territory, informed by municipal- and sub-municipal-level census data. 
Individuals are also assigned to schools and workplaces based on their age and employment status. Unlike many existing models that place individuals into abstract workplaces, our populations match administrative registries of economic establishments like public institutions and companies, and their work-force distributions. Daily commuting flows between municipalities, derived from national mobility surveys and census commuting matrices, are used to ensure consistent individual movements between residential and workplace locations. In this way, we capture the interconnected structure of cities, towns, and rural areas, which is critical for modeling both local and regional diffusion phenomena.
Beyond households and workplaces, we also integrate a rich set of urban points of interest (POI) within each municipality, including restaurants, shops, hospitals, and medical offices. By leveraging national databases of amenities and services, we create a landscape of opportunities that reflects the actual distribution of daily-life environments. Each individual is further assigned a time-use schedule based on national time-use surveys and economic activities business hours. These schedules define how individuals allocate their time across different activities (e.g., work, school, shopping, leisure, healthcare) and locations during the day and week. Together, these elements provide the foundation blocks for building realistic contact networks that span households, educational settings, workplaces, and community spaces.

By combining fine-grained demographic data and commuting flows at census tract level, points of interest and economic activities and behavioral patterns, our synthetic populations represent a major step forward in the construction of geographically explicit individual-based models. 
Compared to existing resources, they offer high spatial resolution at the level of neighborhoods and city blocks. Moreover, cross-country comparability is ensured through a harmonized dataset across Spain, France and Italy. These datasets allows for a broad versatility of applications spanning epidemiology, urban planning, environmental exposure, and social inequality. All steps of the construction process are fully documented, ensuring reproducibility and transparency, and the resulting populations can be readily integrated into individual-based models and statistical analyses.

\section*{Methods}

The generation of a synthetic population relies on the available data sources (see \emph{e.g.}~\cite{chapuis_generation_2022}).
In our case, data sources originate from the statistical institutes of Spain, France, and Italy, and they differ not only in format and spatial/temporal resolution, but also in data-collection methodology, ranging from targeted surveys to census, often referring to diverse metrics and aggregations.
Given this limitation, the three country-specific pipelines exhibit some differences of data ingestion in order to accommodate data sources diversity, yet they share the same logic and the common goal of outputting a consistent format. The resulting dataset encodes critical and, yet, parsimonious information on socio-demographics and behavioral variables that are mostly relevant for epidemic modeling. However, these are further relevant and exploitable for demographic and mobility studies. 
In this context, it is worth mentioning that we have employed OpenStreetMap (OSM) to fetch information on POIs.
The OSM data were spatially integrated with country-specific POIs datasets, such as official schools registries, to produce a harmonized building inventory enriched with attributes such as primary function, floor plan area, geolocation, and usage category.
In this section, we describe preprocessing steps that are common to all three pipelines, as well as the key stages of the synthetic-population generation.
In Table \ref{tab:input_sections}, we have summarized the specific data sources for each country regarding the synthetic generation procedures described in the following subsections.
We refer the reader to the complete description of the data sources and respective references for each nation at the supplementary Tables~\ref{tab:spa_data_sources}, ~\ref{tab:fra_data_sources} and ~\ref{tab:ita_data_sources} for Spain, France and Italy, respectively.


\begin{table*}[p]
\centering
\caption{\textbf{Data inputs.} Country-specific inputs for each phase of synthetic population generation.}
\label{tab:input_sections}
\renewcommand{\arraystretch}{1.4}
\begin{tabular}{|l|l|p{4.5cm}|p{3.5cm}|p{3.5cm}|}
\hline
\textbf{Stage} & \textbf{Country} & \textbf{Description} & \textbf{Resolution} & \textbf{Stratification} \\
\hline

\multirow{12}{*}{\textbf{\shortstack[l]{Households\\\& Dwellings}}}
& \multirow{3}{*}{Spain}   & Age/sex pyramid from census & Municipality & Age, sex \\
\cline{3-5}
&                          & Household microdata & Province \& municipality size & Household type \\
\cline{3-4}
&                          & Cadastral building-level data & Building (cadastral) &  \\
\cline{2-5}
& \multirow{5}{*}{France}  & Age pyramid from census & IRIS & Age, sex \\
\cline{3-5}
&                          & Household type distributions & IRIS & Household type, size, income \\
&                          & Household size               &      & \\
&                          & Household income             &      & \\
&                          & Georeferenced building dataset &    & \\
\cline{2-5}
& \multirow{2}{*}{Italy}   & \multirow{2}{*}{Age/sex pyramid from census} & Census tract & Age (5-year bins), sex \\
\cline{4-5}
&                          &                                                   & Municipality & Age (1-year), sex \\
\cline{3-5}
&                          & Dwelling sizes & Census tract & Dwelling size \\
\cline{3-5}
&                          & Household Microdata & Macro-region &  Age (5-year bins), sex \\
\hline

\multirow{5}{*}{\textbf{\shortstack[l]{Activity\\Assignation}}}
& Spain   & Working-sector distributions from labour-force survey & Province \& municipality & Sex \\
\cline{2-5}
& \multirow{2}{*}{France}  & Activity rates & IRIS & Age \\
\cline{3-5}
&                          & Student share among non-employed & Municipality & Age \\
\cline{2-5}
& \multirow{2}{*}{Italy}  & Professional condition distributions & Municipality & Sex, age, municipality \\
\cline{3-3} \cline{5-5}
&                         & Working-sector distributions &  & Sex, municipality  \\
\hline

\multirow{8}{*}{\textbf{\shortstack[l]{Workplace\\\& School}}}
& \multirow{3}{*}{Spain}   & Cadastral floor-area + OSM   & Building      & Work-sector, province \\
\cline{3-5}
&                          & Employee counts by work-sector      & Province        & Work-sector \\
\cline{3-5}
&                          & OD matrices from mobile-phone data & Municipality & --- \\
\cline{2-5}
& \multirow{2}{*}{France}  & OD matrices   & Municipality & Work-sector, education level \\
\cline{3-3} \cline{5-5}
&                          & Workplace and building sizes &              & Work-sector \\
\cline{2-5}
& \multirow{3}{*}{Italy}   & Census of economic activities  & Census Tract  & Work-sector, employee count, capacity \\
\cline{3-5}
&                          & School census          & Georeferenced & Education level, capacity \\
\cline{3-5}
&                          & OD matrices            & Municipality  & Professional condition (age $>$15) \\
\hline

\multirow{4}{*}{\textbf{\shortstack[l]{Schedule\\Assignation}}}
& Spain   & Working schedules from national Time Use Survey (\emph{Encuesta de Empleo del Tiempo}) & National & Sex, age \\
\cline{2-5}
& France  & Working conditions survey by the Ministry of Labour (DARES): night, weekend and part-time work & National & Socio-professional category \\
\cline{2-5}
& \multirow{2}{*}{Italy}  & Working hours from OSM tags          & Building & Activity type \\
\cline{3-5}
&                         & Collective agreements (trade union reports)     & National & Activity type \\
\hline

\end{tabular}
\end{table*}
\subsection{Household Generation and Dwellings}
For each
census tract
we sample household types from the respective geo-referenced distribution (see \autoref{tab:input_sections} row Household \& Dwelling).
We define a \emph{household type} as a household configuration drawn from the geo-referenced distribution of household compositions. For Italy and Spain, these configurations are described in terms of age and sex of household members (e.g., \emph{one adult male, one adult female, one male child, one teenage female}), while for France they correspond to broader family-situation categories (e.g., \emph{single-man household}, \emph{single-woman with children}  or \emph{couple without children}).
Once the household type is determined, we assign age and sex to each household member, drawing exact ages from the age pyramid of the corresponding municipality or IRIS. Finally, we geo-locate the household in the census tract to match expected population density values.

After this stage each agent has assigned sex, age, household membership, and a dwelling ID encoding both a geographic position and information on co-residence with any other household member.

\subsection{Main activity location assignation}


We draw a professional condition, either \emph{employed}, \emph{student}, or \emph{unoccupied} for all agents, following the corresponding geo-referenced distribution in the census area considered (see \autoref{tab:input_sections} row Activity Assignation). In the case of students, an educational level is assigned based on age and national educational system (i.e., preschool, primary, lower secondary, upper secondary, and university). We will call active agents those agents tagged as employed or students. Employed agents are also assigned an economic sector, depending on the available data, either the sector is drawn first and the destination is restricted to areas hosting workplaces of that sector, or the destination is drawn first and the sector follows from the economic activities located there.
For Italy, the sector is drawn here from the geo-referenced distribution of working sectors; for Spain, it is drawn once the destination is assigned, combining the provincial distribution of workers by sector with the sector-specific buildings available at destination; for France, it is taken from the economic activity of the workplace assigned in the next stage. 



We assign agents to their main activity location based on their conditions: all unoccupied agents have as main activity location their residence area;
for those studying or working, we assign them an activity location taking into account the recurrent origin-destination (OD) mobility flows to estimate the probability of traveling to other census areas.
The sectors of the workers reflect the economic activities of their commute destination; e.g. industry workers commute to areas with industrial facilities; similarly, the destination of students considers the school levels present there, ensuring that no high school student commutes to a municipality with no high schools. 
At the end of this stage, each agent is characterized by a professional condition, among which occupied workers are characterized by a work sector, and students by an education level. Each agent is assigned with an identifier representing the building where they carry out the main activity.

\subsection{Schedule Assignation}

Daily main activity duration are assigned to agents to reproduce realistic temporal patterns.
For Spain, working schedules are drawn from distributions derived from the national Time Use Survey (\emph{Encuesta de Empleo del Tiempo}) of the National Statistical Office INE.
For Italy, working hours are estimated by combining opening-hour information extracted from OSM building tags with typical schedules documented in collective-bargaining agreements published by major trade unions (see~\autoref{tab:ita_working_schedules_1} and~\autoref{tab:ita_working_schedules_2}).
For France, typical work schedules are generated based on national surveys conducted by the French Ministry of Labour, regarding working days and hours across different occupational categories.

Specifically, these datasets provide joint probability distributions over a typical schedules; including night, weekend, and part-time work—conditional on the individual's socio-professional status.

\subsection{Data Records}

The output of the national pipelines is an individual-level dataframe in which each row represents one agent.
\autoref{tab:output_fields} lists the logical fields that are common to all three countries.
The encoding of dwelling IDs and geographic references reflects the respective national data coding: for example, the field \texttt{code} follows INE conventions in Spain, INSEE codes in France, and ISTAT codes in Italy.
An in-depth discussion of each national method, together with the data sources accompanied by links and data access information, is provided in the \hyperref[appendix:methods]{Supplementary Information}.
\begin{table*}[ht]
\centering
\caption{%
  \textbf{Agents' dataset.} This table contains the description of the output fields of the agents' residence location, demographics, job sector and location across Spain, France, and Italy.
}
\label{tab:output_fields}
\small
\begin{tabular}{p{3.2cm} p{1.8cm} p{1.4cm} p{6.8cm}}
\hline
\textbf{Field} & \textbf{Type} & \textbf{Nullable} & \textbf{Description} \\
\hline
\texttt{code}
  & string & No & National municipality identifier (format and coding authority differ by country: ISTAT municipality code for Italy, INE province+municipality ``PPMMM'' code for Spain, INSEE municipality code for France. Finer-grained census-tract/cell-level geography is carried separately in \texttt{cell\_group}/\texttt{cell\_dest}. \\[4pt]
\texttt{dwelling\_id}
  & string & No & Unique dwelling identifier composed of a building reference plus an intra-building sequence number. \\[4pt]
\texttt{household\_type}
& categorical & No & Household typology.
  \begin{enumerate}[start=0, label=\arabic*]
    \item One-person household
    \item Mother/father alone with child
    \item Childless couple
    \item Couple with any children under~25
    \item Couple with all children over~25
    \item Couple with some children under~25 and other people
    \item Childless couple and other people
    \item People who do not form a couple
    \item Other (more than one nucleus)
    \item Collective housing, e.g.\ retirement homes and student residences (France only)
  \end{enumerate} \\[4pt]
\texttt{person\_id}
  & integer & No
  & Person identifier within the household, in the range
    $[1,\,n]$ where $n$ is the total number of co-residents. \\[4pt]
\texttt{age}
  & integer & No & Age of the agent in years. \\[4pt]
\texttt{sex}
  & categorical & No  & Sex of the agent: 0~male, 1~female. \\[4pt]
\texttt{activity}
  & categorical & No & Main activity status:
  0~employed;
  1~unoccupied;
  2~student. \\[4pt]
\texttt{cell\_group}
  & string & No & Spatial region of residence, encoded according to the national grid or census-tract reference system. \\[4pt]
\texttt{sector}
  & categorical & Yes & Economic sector of employment (not assigned to unoccupied agents), taking values in $\{1,2,3\}$. Category definitions are country-specific; see~\autoref{tab:sector_definitions} for national definitions. \\[4pt]
\texttt{cell\_dest}
  & string & No & Spatial region of the main activity location, encoded in the same reference system as \texttt{cell\_group}.
    For unoccupied agents this equals \texttt{cell\_group}. \\[4pt]
\texttt{ref\_dest}
  & string & No & Identifier of the specific building or facility where the main activity takes place (workplace, school); equals the \texttt{reference} field in the buildings dataframe (\autoref{tab:buildings_fields}). \\
\hline
\end{tabular}
\end{table*}

Two companion dataframes complement the agents dataframe above, both keyed to it through \texttt{dwelling\_id}/\texttt{ref\_dest}: a \emph{buildings dataframe} (\autoref{tab:buildings_fields}), giving the geo-referenced inventory of dwellings, workplaces, schools, and other points of interest; and a \emph{schedules dataframe} (\autoref{tab:schedules_fields}), giving each active agent's daily activity timing. 

\section*{Technical Validation} 
\subsection*{National Age Pyramid}
\begin{figure}
    \centering
    \includegraphics[width=\linewidth]{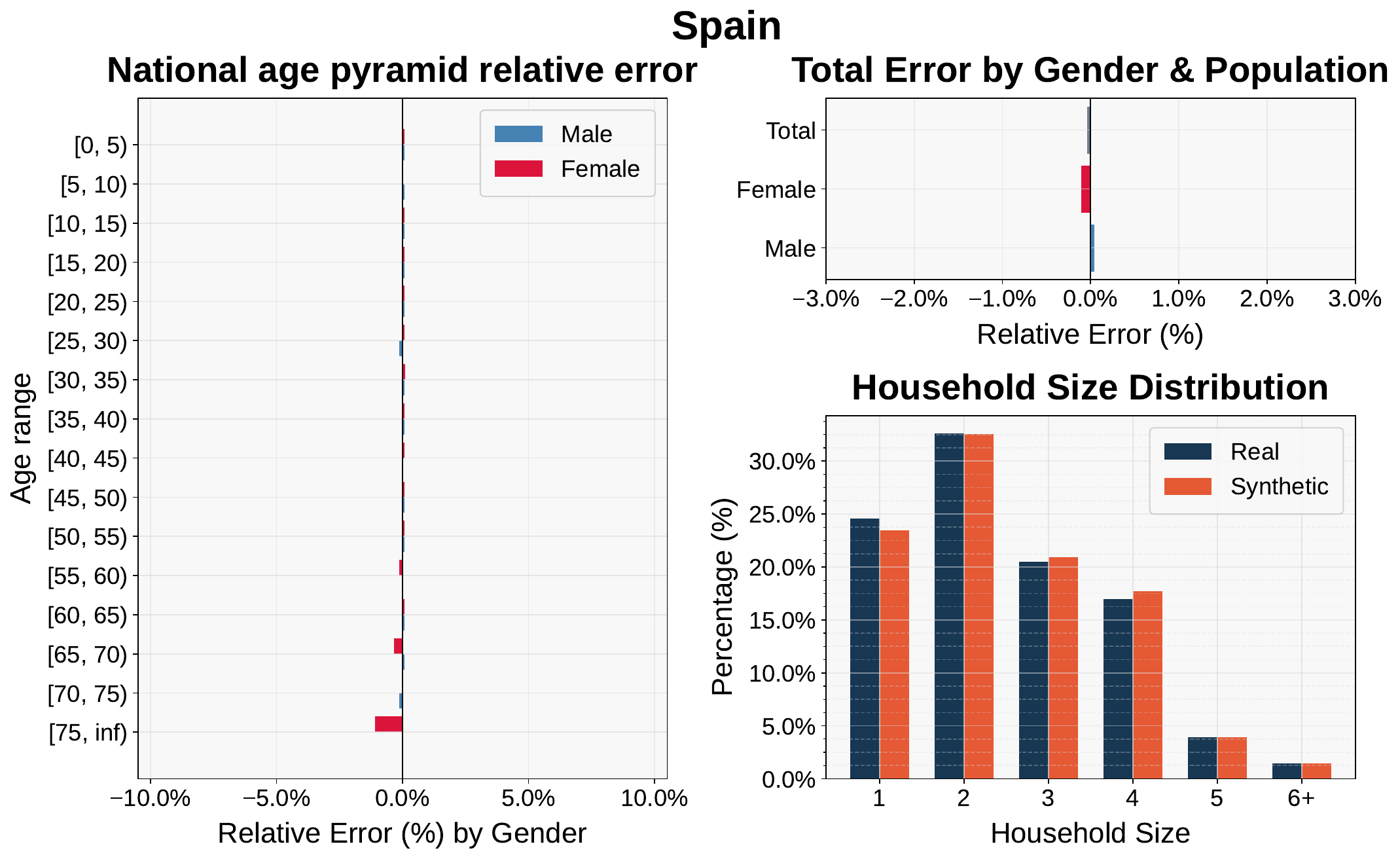}\label{fig:age_pyramid_spa}
    
    \vspace{0.1cm}
    
    \includegraphics[width=\linewidth]{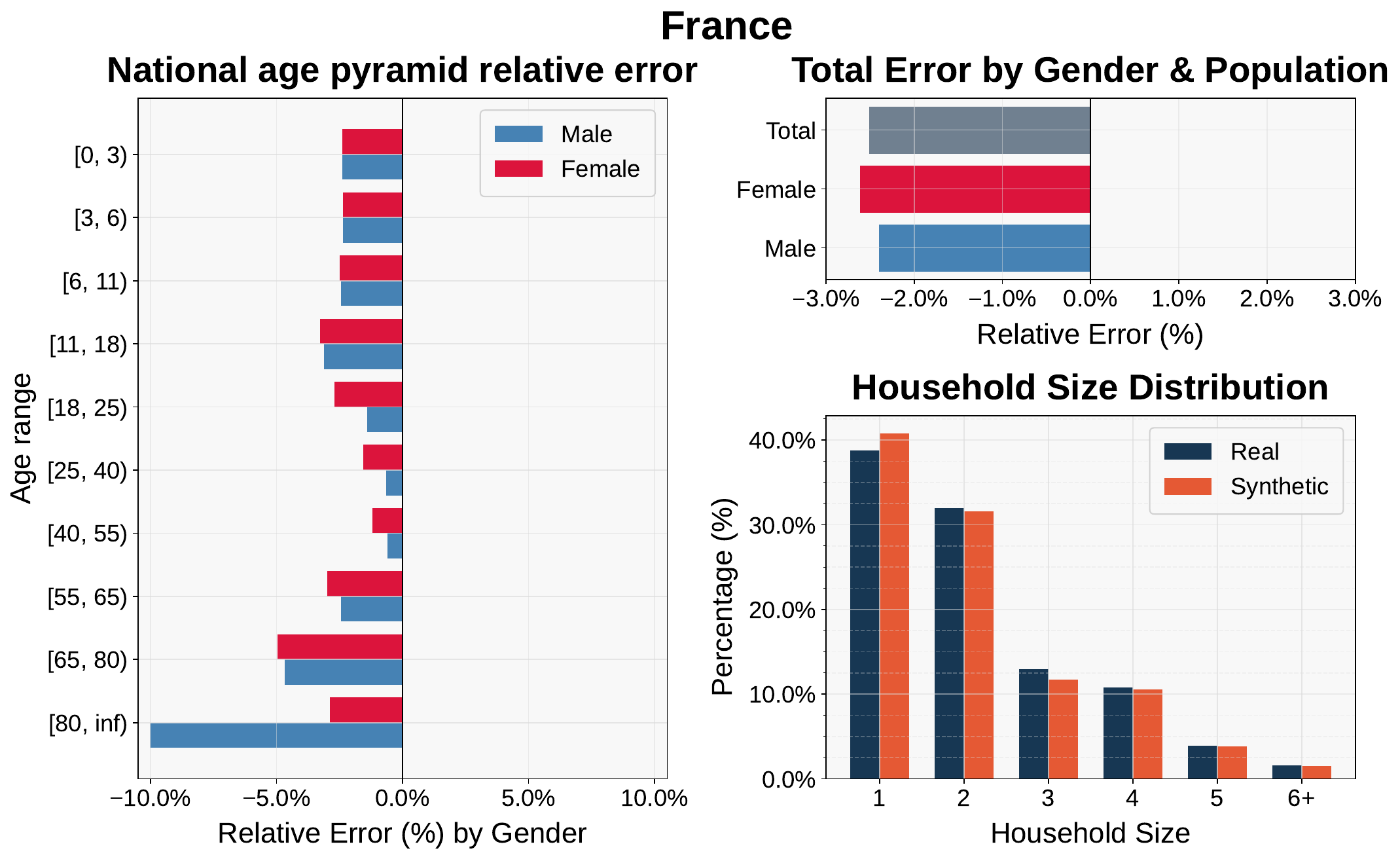}\label{fig:age_pyramid_fra}
    
    \vspace{0.1cm}

    \includegraphics[width=\linewidth]{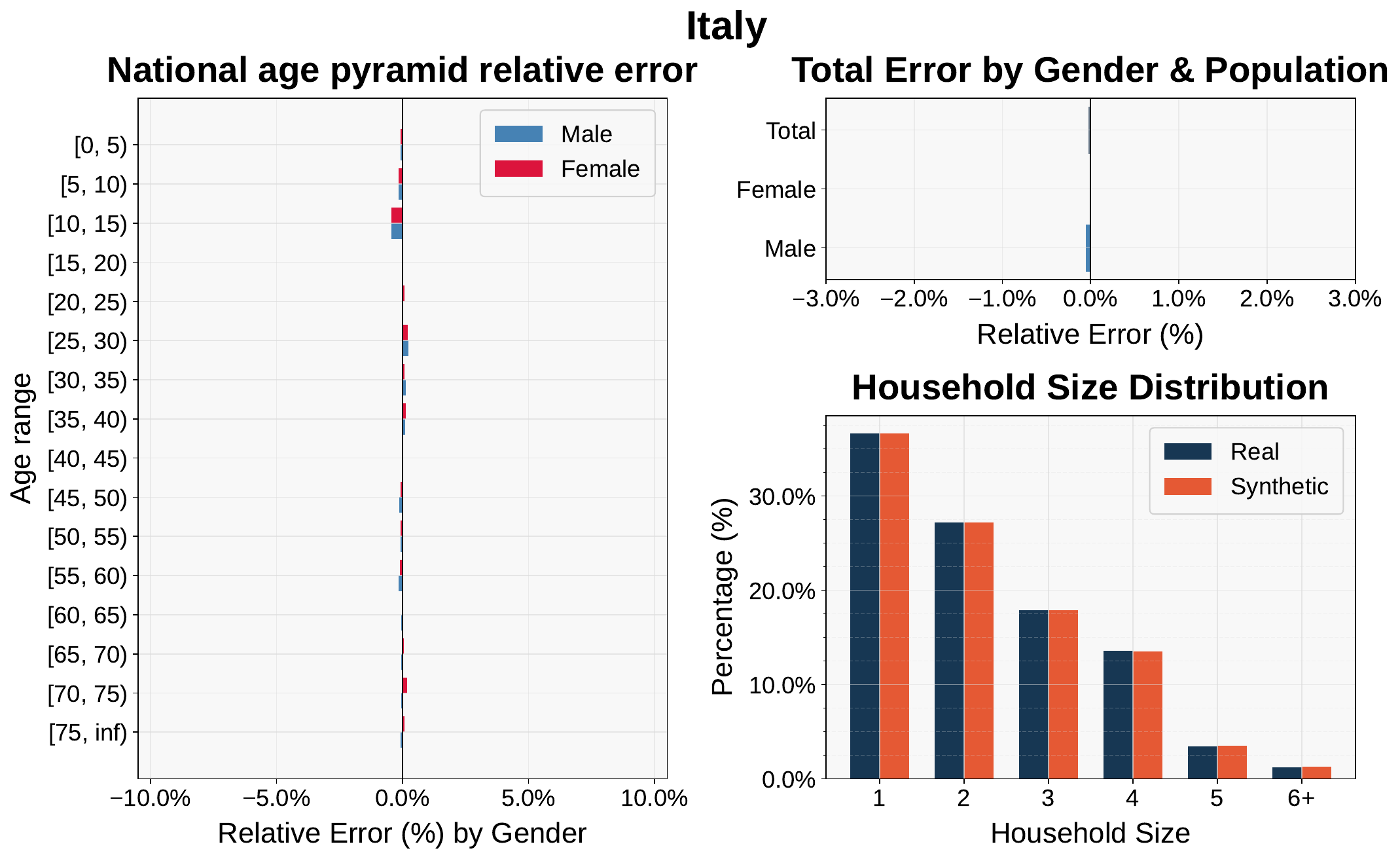}\label{fig:age_pyramid_ita}

    \caption{\textbf{Comparison of synthetic and ground truth age distributions.} Left: Population pyramid showing relative errors by age group and gender. Right: Total population errors aggregated by gender and overall population and distribution of households by size.
    Age groups are displayed in 5-year intervals for Italy and Spain, with the exception of the final 75+ category. For France, age groups follow the non-uniform intervals defined in the ground truth census data.
    }\label{fig:age_pyramid}
    
\end{figure}

We validate the age distribution of the synthetic population in~\autoref{fig:age_pyramid}, by comparing the demographic structure against ground truth data aggregated at the national level. The population pyramid displays relative errors between synthetic and real populations across age groups and genders.
Age groups displayed in 5-year intervals for Italy and Spain, with the exception of France, where age groups follow the non-uniform intervals defined in the respective ground truth ( \autoref{tab:fra_data_sources}, \textit{Data Source 15}), namely: $[0, 3)$, $[3, 6)$, $[6, 11)$, $[11, 18)$, $[18, 25)$, $[25, 40)$, $[40, 55)$, $[55, 65)$, $[65, 80)$, and $80+$. 

The left panel in ~\autoref{fig:age_pyramid} shows the age-specific relative errors by gender, where each horizontal bar represents the percentage deviation of the synthetic population from ground truth for males (blue) and females (red). The right panel presents the overall population totals by gender and for the entire population.

\paragraph*{Spain}
The synthetic population has an overall relative error of $-0.04\%$ on the entire population, corresponding to $\sim 1.8 \times 10^4$ fewer individuals. With regard to gender-specific errors, males show a deviation of $+0.04\%$ and female $-0.11\%$.
The Mean Absolute Percentage Error (MAPE) is $0.12\%$ across all age groups ($0.08\%$ for males and $0.16\%$ for females).

\paragraph*{France}
The synthetic population covers metropolitan France only; the overseas departments are therefore excluded from the validation.
The synthetic population has an overall relative error of $-2.51\%$ on the entire population, corresponding to $\sim 1.6 \times 10^6$fewer individuals.
With regard to gender-specific errors, males show a deviation of $-2.40\%$ and females $-2.62\%$.
The MAPE is $2.85\%$ across all age groups ($3.01\%$ for males and $2.68\%$ for females).

\paragraph*{Italy}
The synthetic population has an overall relative error of $-0.03\%$ on the entire population, corresponding to $\sim 1.6 \times 10^4$ fewer individuals.
With regard to gender-specific errors, males show a $-0.05\%$ deviation and females a negligible $-0.002\%$ deviation. 
The MAPE is $0.11\%$ across all age groups ($0.11\%$ for males and $0.11\%$ for females).
The [10-15) age group exhibits the most pronounced deviation among all age ranges, with the synthetic population showing a deviation of $0.45\%$ for females.

\subsection*{Household Size Distribution}

The synthetic population replicates the ground truth household size distribution closely for Spain and Italy, and reasonably well for France (\autoref{fig:age_pyramid}).
For Spain, the largest deviation is $-1.10$ percentage points for single-person households, offset mainly by an over-representation of four-person ($+0.75$ percentage points) and three-person households ($+0.46$ percentage points), while all other sizes deviate by at most $0.11$ percentage points.
For France, the largest deviation is $+2.02$ percentage points for single-person households, offset mainly by an under-representation of three-person households ($-1.22$ percentage points), while all other sizes deviate by less than $0.4$ percentage points.
The total number of French households is underestimated by $7.94\%$.
For Italy, the share of households of each size deviates by at most $0.04$ percentage points from the ground truth (households of size $6+$), and the total number of households is overestimated by $0.33\%$.


\subsection*{Municipality-level Population Distribution}

\begin{figure}
    \centering
    \includegraphics[width=\linewidth]{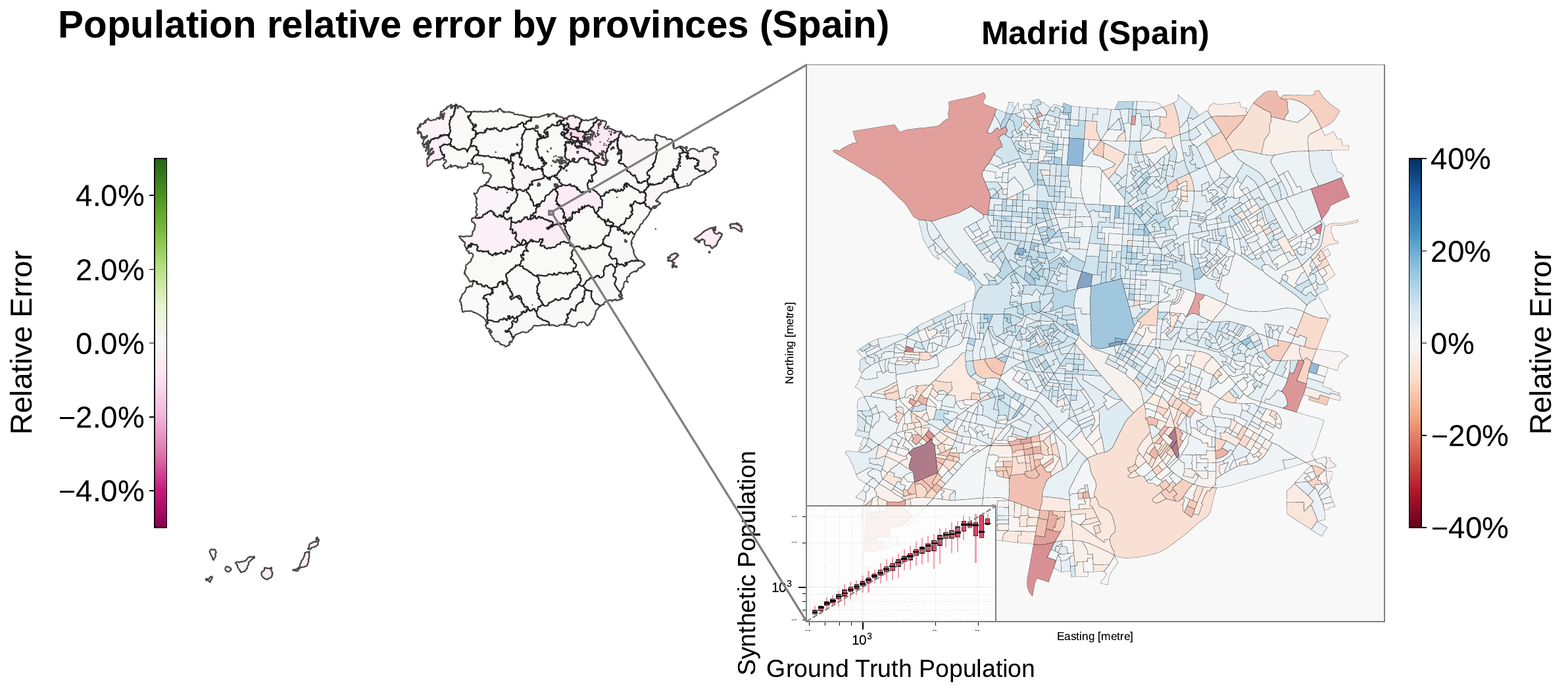} 
    
    \vspace{0.1cm}

    \includegraphics[width=\linewidth]{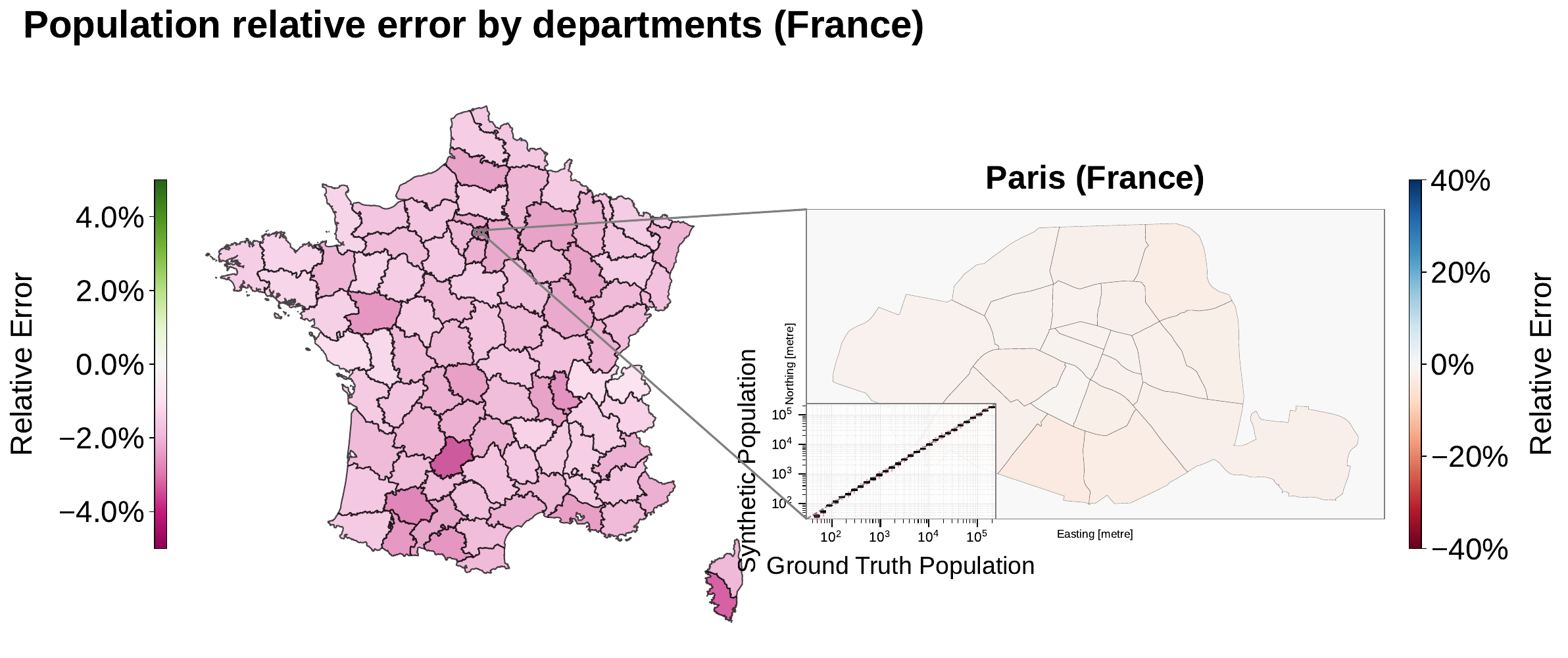}
    
    \vspace{0.1cm}

    \includegraphics[width=\linewidth]{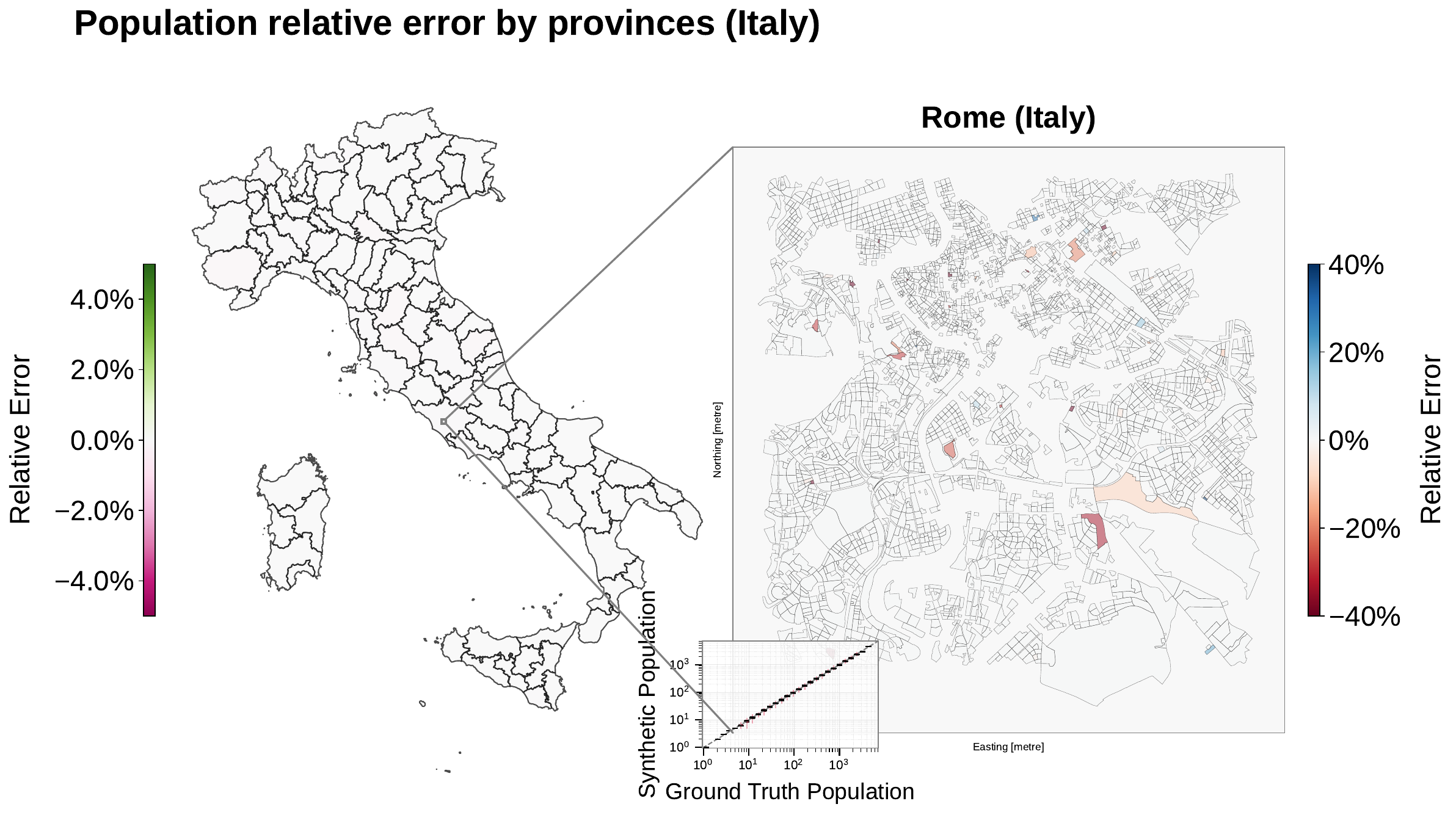}

    \caption{\textbf{Spatial validation of synthetic population at NUTS-3 and LAU levels}. Maps show relative errors between synthetic and ground truth population counts, with red shades indicating underestimation and blue shades indicating overestimation. Light colors across most territories indicate low relative errors. On the left, national maps showing relative errors at NUTS-3 (province/department) level; on the right, capital insights showing relative errors at LAU (census tract) levels.}
    \label{fig:municipality_error_map}
\end{figure}

We validate the spatial distribution of the synthetic population in~\autoref{fig:municipality_error_map} by comparing municipality-level synthetic population counts against ground truth census data.
The maps display relative errors at Local Administrative Units (LAU) level of Rome, Paris and Madrid, with color coding representing the relative error from red ($-40\%$) to blue ($+40\%$), with zero coded in white for perfect agreement ($0\%$).

\paragraph*{Spain}
The spatial agreement with ground truth data is heterogeneous across the capital's census tracts. Population counts tend to be slightly overestimated in several central and northern neighborhoods, while both over- and underestimated counts are visible in peripheral areas: over the $1{,}597$ census tracts shown in the map, the median relative error is $+5.9\%$, with $90\%$ of the tracts between $-13.7\%$ and $+16.5\%$. At the national scale, the NUTS-3 map shows how the synthetic population closely matches total population counts across Spanish provinces, including the islands, with relative errors ranging from $-1.5\%$ to $0.0\%$ (median $-0.05\%$).

\paragraph*{France}
The synthetic population shows strong spatial agreement with ground truth data across the census tracts (IRIS) of the capital. Tract-level relative errors are small and mostly negative, with a median of $-2.6\%$ and $90\%$ of the tracts between $-10.1\%$ and $+0.1\%$.
The national map shows the NUTS-3 level (departments) relative errors: all metropolitan departments are slightly underestimated, with errors ranging from $-5.8\%$ to $-1.2\%$ (median $-2.3\%$).

\paragraph*{Italy}
The synthetic population shows strong spatial agreement with ground truth data across most census tracts of the municipality of Rome.
The figure shows that the relative error in Rome is narrowly concentrated around zero (white, perfect agreement): around $98.4\%$ of the $2{,}798$ census tracts shown in the map are reproduced exactly. Population counts deviate only in a few tracts (with relative errors between $-100\%$ and $+37.5\%$), mainly because of the constraints of the census tract statistics (e.g., no valid household can be formed in a census tract whose residents are only children).
The national map shows how the population allocation methodology scales well to the NUTS-3 level, with all provinces exhibiting a good agreement with the ground-truth values: relative errors range from $-0.13\%$ to $-0.002\%$ (median $-0.03\%$).

\subsection*{Origin Destination Flows}
\begin{figure}
    \centering
    \includegraphics[width=0.8\linewidth]{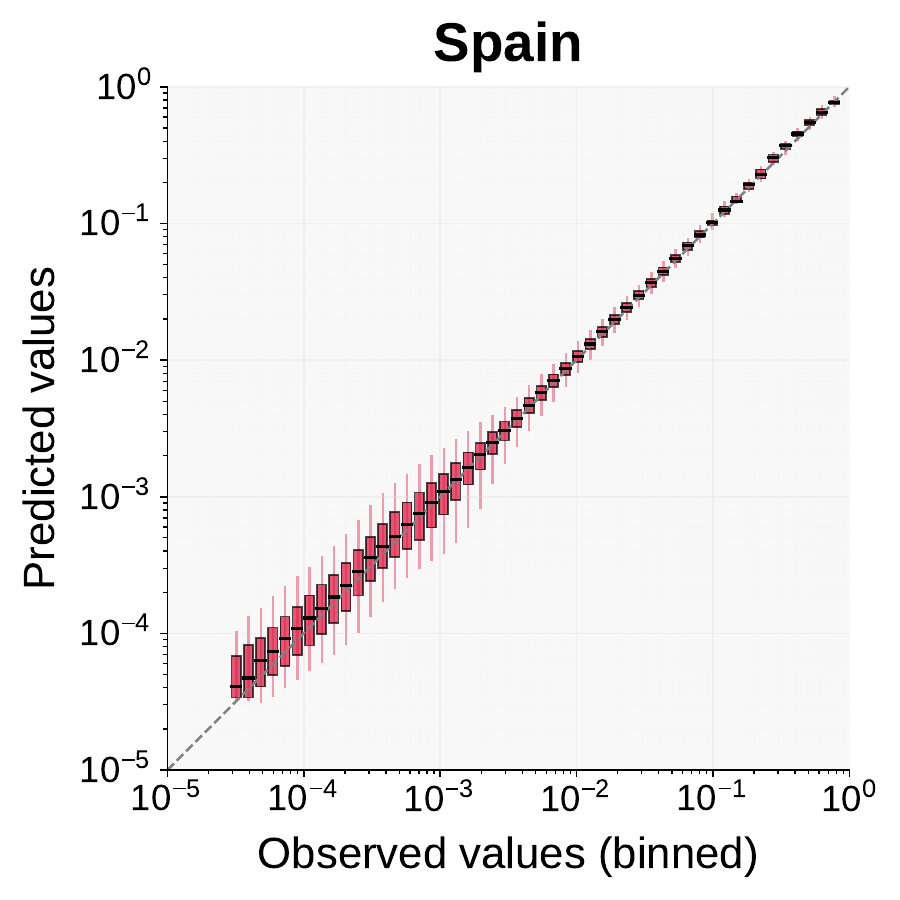}
    \label{fig:od_flow_comparison_spa}
    \vspace{0.1cm}
    
    \includegraphics[width=0.8\linewidth]{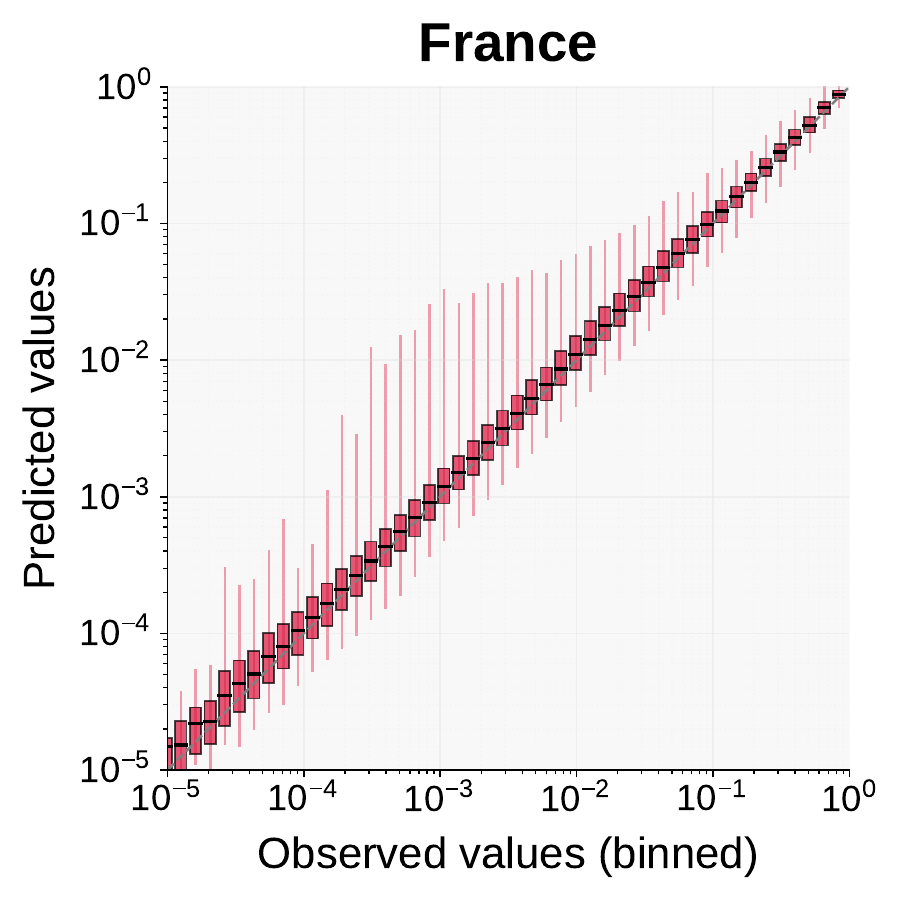}
    \label{fig:od_flow_comparison_fra}
    \vspace{0.1cm}

    \includegraphics[width=0.8\linewidth]{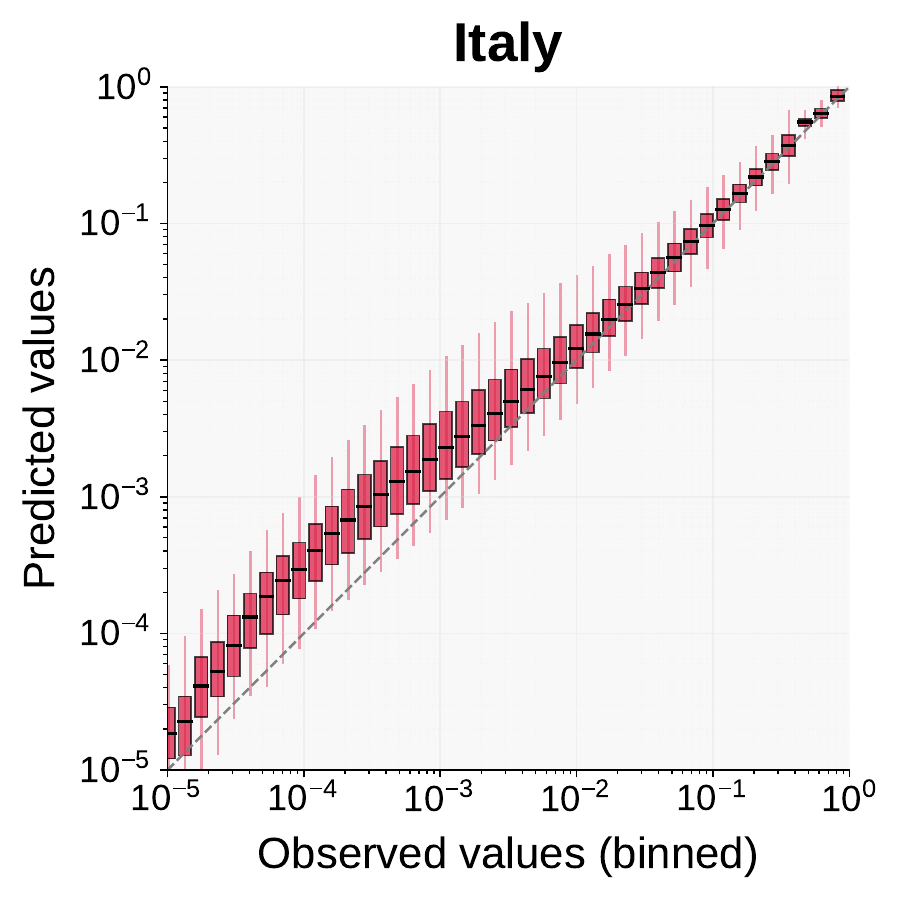}
    \label{fig:od_flow_comparison_ita}

    \caption{\textbf{Comparison of synthetic and ground truth origin-destination (OD) travel probabilities for commuting mobility}. Each entry represents an origin-destination couple, plotting synthetic data frequency (y-axis) against ground truth travel probability (x-axis).
    }
    \label{fig:od_flow_comparison}
\end{figure}

We validate the Origin Destination (OD) commuting flows, considering students traveling to educational facilities and workers traveling to workplaces in~\autoref{fig:od_flow_comparison}. We compare synthetic population commuting mobility patterns against ground truth data from the three national censuses. As a unit of measure we choose the OD travel probabilities $p_{i,j}$, as the critical distribution shaping diffusion phenomena, e.g. epidemics, and population level estimates of hazards exposure. For sake of visualization we binned the expected values and showed the synthetic values in boxplots, in order to highlight the variation on the synthetic values on the y-axis with respect to the ground truth expected travel probabilities on the x-axis. Overall the boxplots show a good agreement across all trip routes, especially on the primary ones, i.e. higher $p_{ij}$ on the top right corner of the boxplot, without neglecting accuracy for the low connectivity routes (bottom left of the boxplots.) The three countries show Pearson correlation coefficients of $r = 0.9998$, $r = 0.9612$, and $r = 0.9842$ for Spain, France, and Italy, respectively (all $p < 0.001$).

\subsection*{Co-presence matrices in households}

\begin{figure}
    \centering
    
        \noindent
    \begin{minipage}[c]{0.75\linewidth}
        \includegraphics[width=\linewidth]{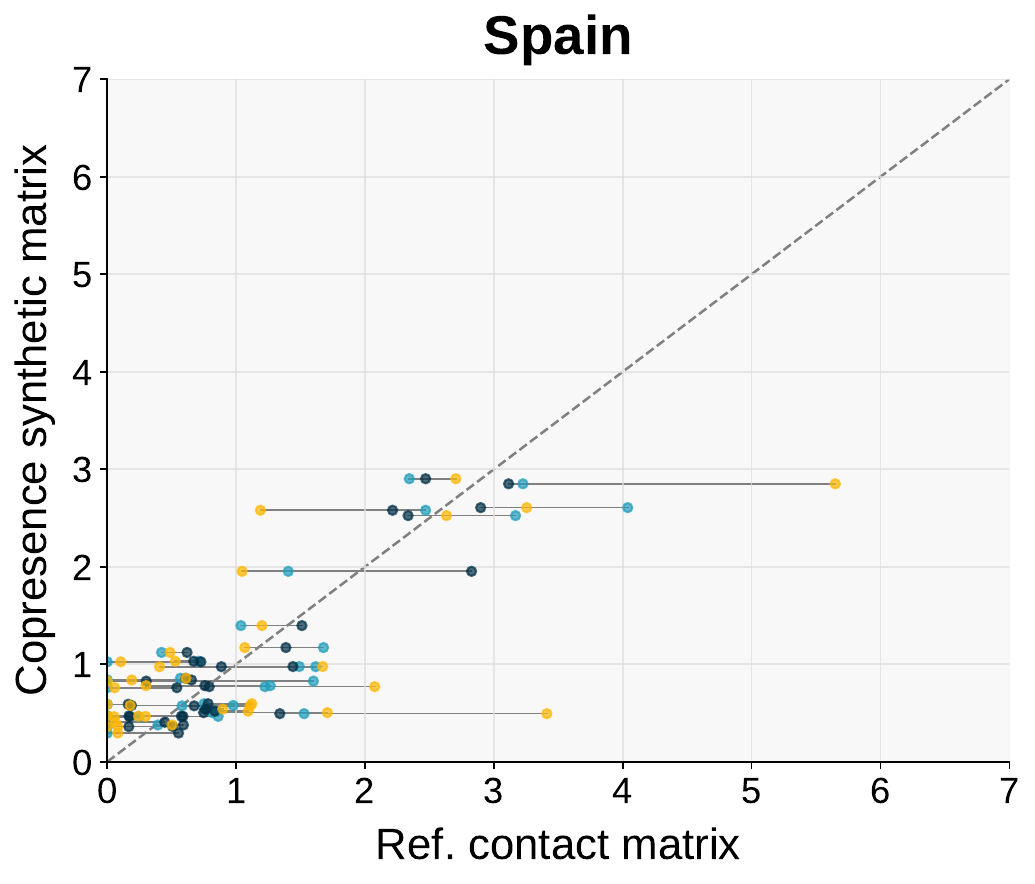}
        \label{fig:copresence_comparison_spa}
    \end{minipage}%
    \hfill
    \begin{minipage}[c]{0.18\linewidth}
    \end{minipage}

    \vspace{-0.3cm}
    
    \noindent
    \begin{minipage}[c]{0.75\linewidth}
        \includegraphics[width=\linewidth]{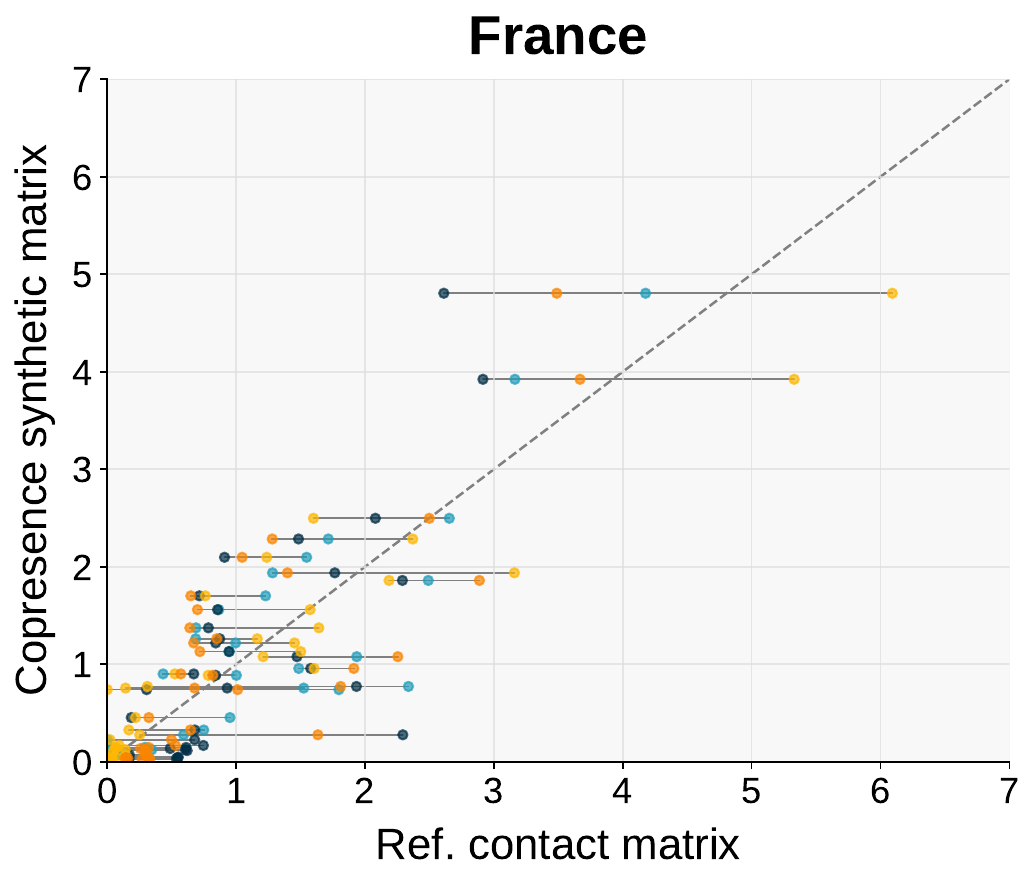}
        \label{fig:copresence_comparison_fra}
    \end{minipage}%
    \begin{minipage}[c]{0.25\linewidth}
        \centering
        \includegraphics[width=1.2\linewidth]{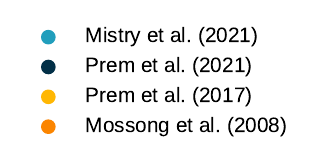}
    \end{minipage}
    
    \vspace{-0.3cm}
    
    \noindent
    \begin{minipage}[c]{0.75\linewidth}
        \includegraphics[width=\linewidth]{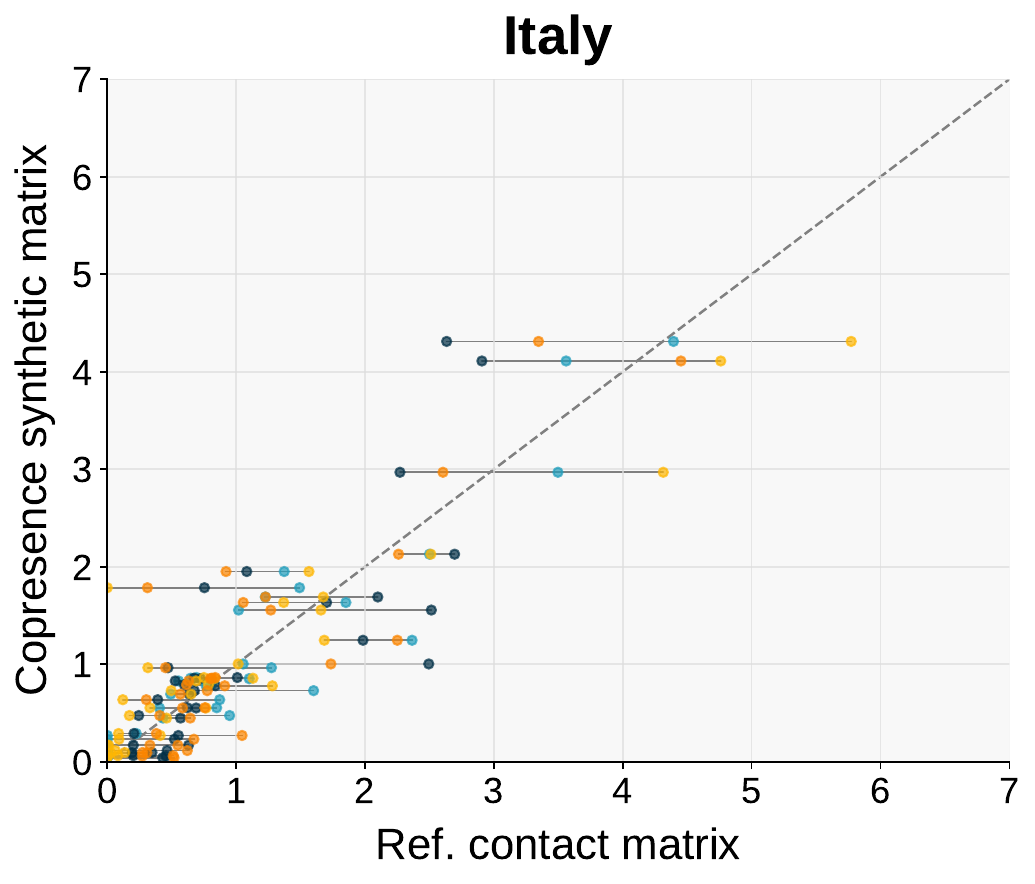}
        \label{fig:copresence_comparison_ita}
    \end{minipage}%
    \hfill
    \begin{minipage}[c]{0.18\linewidth}
    \end{minipage}
    
    \caption{\textbf{Household co-presence patterns validation.}
    Each panel represents the results from each Country.
    Points from the same matrix element across different empirical
    references are connected by horizontal lines.
    The dashed diagonal represents perfect
    agreement between synthetic and empirical data. All matrices were
    normalized by their mean. ~\autoref{tab:contact_matrix_validation} summarize the correlation score between
    the synthetic household co-presence structure and the data from each study.}
    \label{fig:contact_matrix_validation}
\end{figure}

To validate the household composition of the synthetic population, we computed a co-presence matrix capturing the average number of individuals of age group $j$ sharing a household with an individual of age group $i$, using five age classes: 0–9, 10–17, 18–29, 30–59, and 60-79. 
For sake of generalizability, we did not impose any contact rule on the co-presence matrix, hence assuming that co-presence directly implies contacts in household.
We compared our co-presence matrices with empirical daily contact rates extracted from two state-of-the-art national data collections and two studies reconstructing synthetic contact matrices from public data.
Namely, we report comparisons with Polymod~\cite{mossong2008polymod} for Italy and France, and Prem. et al. \cite{prem2017projecting,prem2021projecting} and Mistry et al.~\cite{mistry2021inferring} for all the three countries.
To enable a meaningful comparison in terms of household mixing structure, each matrix was normalized by its own mean, yielding a rescaled dimensionless matrix reflecting the relative structure of within-household mixing between age groups.
This allows a direct visual assessment of how well the synthetic population reproduces the age-assortative structure observed in reference surveys and state-of-the-art synthetic household contact data. In the context of their main use in epidemic modeling, contact matrices are invariant to scalar normalizations, as any scalar would be absorbed in the estimated transmission rate during model calibration.
In~\autoref{fig:contact_matrix_validation} we show the normalized synthetic matrices on the y-axis against each of the four normalized empirical matrices on the x-axis, where every dot encodes an age-by-age mixing rate.
As shown in the Figure, the state-of-the-art reference studies show high variability for most of the mixing rates, proving that, even in absence of any contact rule which can be imposed on top of the synthetic households, our co-presence structures are in line with reference studies.
At a quantitative level, Table~\ref{tab:contact_matrix_validation} reports Pearson's $r$ and Spearman's $\rho$ correlation coefficients between the synthetic co-presence matrix and each reference dataset.
For Spain, the highest agreement is observed with Prem et al.\ (2021)~\cite{prem2021projecting} ($r = 0.91$, $\rho = 0.73$), while for France the best performance is achieved with Prem et al.\ (2017)~\cite{prem2017projecting} ($r = 0.94$, $\rho = 0.91$). For Italy, the highest agreement is found with Mistry et al.\ (2021)~\cite{mistry2021inferring} ($r = 0.94$, $\rho = 0.90$), closely followed by Prem et al.\ (2017)~\cite{prem2017projecting} ($r = 0.94$, $\rho = 0.83$).

\begin{table}[ht!]
\centering
\renewcommand{\arraystretch}{1.4}
\caption{\textbf{Comparison of synthetic household co-presence patterns with reference studies.} The table shows Pearson's and Spearman's correlation coefficients for the synthetic population household co-presence matrix with respect to reference studies on contact patterns.}
\label{tab:contact_matrix_validation}
\begin{tabular}{|l|l|c|c|}
\hline
\textbf{Nation} & \textbf{Dataset} & \textbf{Pearson's $r$} & \textbf{Spearman's $\rho$} \\
\hline
\multirow{3}{*}{Spain} & Mistry et al. (2021) & 0.84 & 0.67 \\
\cline{2-4}
 & Prem et al. (2021) & 0.91 & 0.73 \\
\cline{2-4}
 & Prem et al. (2017) & 0.67 & 0.56 \\
\hline
\multirow{4}{*}{France} & Mistry et al. (2021) & 0.88 & 0.84 \\
\cline{2-4}
 & Prem et al. (2021) & 0.75 & 0.81 \\
\cline{2-4}
 & Prem et al. (2017) & 0.94 & 0.91 \\
\cline{2-4}
 & Mossong et al. (2008) & 0.82 & 0.84 \\
\hline
\multirow{4}{*}{Italy} & Mistry et al. (2021) & 0.94 & 0.88 \\
\cline{2-4}
 & Prem et al. (2021) & 0.79 & 0.85 \\
\cline{2-4}
 & Prem et al. (2017) & 0.97 & 0.86 \\
\cline{2-4}
 & Mossong et al. (2008) & 0.89 & 0.73 \\
 \hline

\end{tabular}
\end{table}





\section*{Discussion}

In this work, we introduced a novel spatially explicit synthetic population for Spain, France, and Italy, endowed with demographic, socioeconomic, and commuting behavioral attributes. Our work advances the state of the art in spatially explicit synthetic population generation for Spain, France and Italy in several key dimensions. The primary contribution of this work is the dataset itself: a geo-referenced population of digital agents that can serve as a shared resource for a broad range of simulation-based and analytical studies.
In all cases, the populations demonstrated a satisfactory level of agreement with the reference data. We performed several validations across the three countries, including internal validation on demographic marginals and commuting flows from census data. Moreover, we performed external validation on household co-presence against state-of-the-art contact surveys and spatially-agnostic synthetic contact matrix reconstructions. The comparisons for Italy and France also exhibit a good agreement with the Polymod survey \cite{mossong2008polymod} ($r = 0.87$ and $r = 0.82$, respectively), proving that even without imposing a contact rule, synthetic household co-presence reconstruction shows a high fidelity to expected levels of age-structured household mixing in both countries. Overall, the good agreement with studies like Mistry et al. \cite{mistry2021inferring} in the three countries show that our main contribution, i.e. adding geographical explicitness to synthetic populations, does not undermine fidelity to socio-demographic aspects like age-stratified household mixing achieved from state-of-the-art studies that neglect the geographic dimension.
This is also reflected in the good agreement with expected travel probabilities for commuting mobility, proving high accuracy on both primary routes, i.e. showing the highest probability from any location $i$ to destinations $j$, and lowest connectivity routes, i.e. the routes that are most rarely used from commuters. This is especially important for modeling diffusion phenomena with accuracy, as in the case of epidemic spatial transmission models, where the travel probabilities shape the arrival times and the intensity of the spreading epidemic at destination \cite{balcan2009multiscale, tizzoni2014use, pullano2025characterizing, mazzoli2021interplay}. The agreement in terms of spatial-connectivity of the synthetic mobility network, combined with that on social mixing, allows to use our synthetic populations for informing multi-scale models combining realistic mobility and mixing patterns, which can be relevant for applications in urban studies \cite{xu2025using}, epidemic models \cite{cota2021infectious} and demography \cite{predhumeau_synthetic_2023}. Most importantly, socio-demographic aspects have proven critical in shaping the adherence in response to public health interventions \cite{valdano2021highlighting, pullano2020evaluating, gauvin2021socio}, vulnerability to infection \cite{davies2020age} and severe outcomes \cite{verity2020estimates}. Our data allows to account for individual socio-demographic information that is critical for epidemic inequalities estimates, informing models with more realistic heterogeneity in infection susceptibility, vaccine adoption and response to policy, without losing the spatial fidelity of state-of-the-art meta-population models and demographic insights of age-structured models. 
Beyond socio-demographic stratification, our data allows to dive more on recent advances linking epidemic intensity to local urban structures \cite{hazarie2021interplay, aguilar2022impact}, opening the door to testing the impact of settings-specific and temporal interventions on the transmission of pathogens \cite{colosi2023minimising, smith2022evaluating, di2021impact}.

For Spain, previous work has focused mainly on city-scale digital twins for epidemiological applications. In particular, ~\cite{rodriguez2025multilayer} generated detailed synthetic populations for five Spanish cities by combining census and mobility data to reconstruct multilayer contact networks to model infectious diseases. In contrast, our work provides a nationwide synthetic population covering the entire Spanish territory, integrating cadastral information with census and official statistical data to assign households to real residential buildings across both urban and rural areas. This substantially extends the spatial scope from a limited set of metropolitan areas to the national scale while increasing the spatial resolution of the residential environment. Moreover, whereas the previous work was specifically designed to support epidemic simulations, our dataset is conceived as a general-purpose spatial infrastructure.

For France, several synthetic populations have been produced, each tied to a specific application. Hörl et al. ~\cite{horl2021synthetic} developped a reproducible open-data pipeline generating households, persons, and daily activity chains for the Paris–Île-de-France region, discretizing residence, workplace, and school to real addresses for agent-based transport simulation. the OLYMPUS-POPGEN project ~\cite{elessa_etuman_olympus-popgen_2024} generated a dwelling-level population for the same region using a Monte Carlo Markov Chain approach, allocating households to housing by income and price class, to assess exposure to air quality. Moreno et al ~\cite{moreno_lopez_anatomy_2021} built an agent-based population of Metropolitan France from INSEE census data, assigning individuals to a municipality, household, and workplace or school with inter-municipality commuting flows, as the basis of a multi-setting contact network for epidemic modelling. In contrast, our work provides a single nationwide synthetic population, spanning urban and rural areas rather than a single region, in which every individual is located in a real building through the integration of building and equipment databases—including. Beyond ordinary households, we implemented the population living in collective establishments such as retirement homes and student residences. Combined with activity status and home–work and home–school assignment, this makes the dataset a general-purpose spatial infrastructure, reusable across various types of studies.

For Italy, two notable prior contributions exist: Little Italy (Iozzi et al.~\cite{iozzi_little_2010}), which generated a synthetic population of ~10,000 agents based solely on time-use data without explicit geographic assignment, and the model by Cioffi Degli Atti et al.~\cite{ciofi_mitigation_2008}, which stratified the population at the municipality, province, and regional level using 2001 census data. Compared to these works, our population is georeferenced at the census tract level, a substantially finer spatial granularity, and integrates origin-destination commuting matrices, time-use schedules, and points of interest (POI) to assign each agent a specific activity, location, and daily schedule. 
Under the point of view of the generation of multiple countries' synthetic populations, Fumanelli et al. ~\cite{fumanelli_inferring_2012} created the synthetic population of 26 European countries.
While they manage to create a comprehensive cross-country population with harmonized data structures, the buildings created for the activities are generated out of national-level size distribution, while in our case, we collected data directly from OSM, national Ministries of Education, or from the census tract data, providing data on capacity and geographical reference.   
Despite the overall satisfactory performance of the population synthesis pipeline, a number of limitations should be acknowledged.

For Spain, several limitations stem from the availability and granularity of the underlying data sources rather than from the population synthesis methodology itself. First, many of the datasets used are only available in aggregated form, inevitably limiting the spatial and behavioral resolution that can be achieved. For example, the Spanish Time Use Survey (INE) provides representative activity patterns but only through aggregated statistics, preventing the reconstruction of detailed individual behavioral trajectories. Similarly, the household microdata available for population synthesis are based on relatively limited survey samples, which may underrepresent uncommon household configurations and introduce uncertainty when reproducing local demographic heterogeneity.

A similar limitation applies to mobility data. Mobile phone-based origin-destination matrices provide extensive spatial coverage but describe aggregated mobility flows without information on the purpose of the trip or the activity performed at each destination. While more detailed travel surveys including trip purposes exist for specific metropolitan areas (e.g., the Madrid Region Mobility Survey \cite{edm2018}), no equivalent nationwide dataset is currently available, requiring assumptions when assigning activities and destinations across the country. Finally, although cadastral information and OpenStreetMap constitute complementary and highly valuable sources for locating residential buildings and activity places, both datasets present incomplete coverage and varying levels of thematic detail. In particular, cadastral records often classify buildings into broad functional categories, while the completeness of OpenStreetMap depends on local mapping efforts, potentially affecting the assignment of activities and points of interest in some areas.

For Italy, household generation relies on microdata from the 2001 census survey, while origin-destination matrices are derived from the 2011 census. Although we believe the methodology remains valid and will be applicable to more recent data when it becomes available, these temporal discrepancies may limit the accuracy of the generated population in capturing current demographic and mobility patterns.
Furthermore, origin-destination matrices show high variability in smaller municipalities, where the sparsity of agents induces instability in the estimated commuting flows. This effect should be taken into account when using the synthetic population for fine-grained mobility analyses in low-density areas.
The generated synthetic populations open a wide range of possibilities for future research and applied work, including individual-based modeling of infectious disease dynamics, exposure-to-hazard assessments for environmental and health risk analyses, transportation and mobility simulations, and broader investigations in demography, urban planning, and social sciences.

For France, several limitations arise from data temporalities, spatial aggregation, and facility capacity estimates.
First, while most input datasets date from after 2021, not all sources share the same reference year. This minor temporal mismatch may introduce slight inconsistencies when cross-referencing demographic distributions, employment data, and spatial infrastructures, slightly altering the alignment between residential patterns and local economic activities.

Second, commuting movements are derived from inter-communal origin-destination matrices that lack stratification by job type or socio-economic status. Consequently, destination assignments reflect general spatial commuting patterns rather than occupation-specific spatial behaviors, which might smooth out disparities in mobility ranges across socio-professional categories.
Third, due to the lack of exhaustive national data on individual school enrollments, building capacities were estimated proportionally from their footprint area. While this proxy reliably scales with facility size, it may cause local discrepancies in student-to-building assignments in cases of highly heterogeneous building usage or space optimization.
Finally, origin-destination matrices show sparse flows for small municipalities, where few observed commuters induce high variance in the estimated destination probabilities. This instability should be taken into account when using the synthetic population for fine-grained mobility analyses in low-density areas, consistent with the limitation reported for the Italian case.

Overall, our approach is generalizable to further countries where census data on population density, household configuration, commuting and OSM POIs are available. 
Our work opens the doors to a further level of refinement of state-of-the-art synthetic populations generation, allowing a plethora of applications ranging from computational epidemiology, urban studies, mobility modeling and demographic analyses with heightened spatial granularity, high fidelity with respect to expected statistics, and wide applicability.

\section*{Data Availability Statement}
The synthetic populations are available in the DIGITAL.CSIC repository at \url{https://doi.org/10.20350/digitalCSIC/17518}, \textbf{please notice that data will be available upon publication}. Input data are available from INE, INSEE and ISTAT, as detailed in the Supplementary Information.

\section*{Code Availability Statement}
The code to reproduce the main figures will be made publicly available upon publication.

\section*{Acknowledgements}
MA, AT and JJR acknowledge partial financial support from the Spanish State Research Agency
(MICIU/AEI/10.13039/501100011033) and FEDER (UE) under project COSASTI ( PID2024-157493NB-C22), and the Mar{\'\i}a de Maeztu project CEX2021-001164-M. DT, JDI, MM and DP acknowledge support from the Lagrange Project of the ISI Foundation, funded by Fondazione CRT.
EV acknowledges partial financial support from the ANR (Agence Nationale de la Recherche) JCJC grant DiscoReel (ANR-25-CE45-5346-01) and from the ANRS Maladies Infectieuses Émergentes grant PP-IMPACT (ANRS00839-R-PR).
MA, DT, JDI, AT, MM, DP, JJR and EV  acknowledge support from the EU Commission in the framework of the EU Horizon Project EOSC-SIESTA (101131957).
DT wishes to thank Gabriele Torre for assisting him in the writing of the paper and the creation of the images in the validation section.

\section*{Author Contribution}
Conceptualization: MM, DP, JJR, EV.
Writing—original draft: DT, MA, BF, MM, JJR.
Methodology: DT, MA, BF, RA, AT, MM, JJR.
Formal analysis: DT, MA, BF, RA, JDI.
Software: DT, MA, BF, AT.
Visualization: DT, MM.
Investigation: DT, MA, BF, JDI.
Data curation: DT, MA, BF, RA, JDI.
Project administration: MM, DP, JJR, EV. 
Writing—review and editing: All authors.
Supervision: MM, DP, JJR, EV. 
Funding acquisition: MM, DP, JJR, EV.

\bibliographystyle{nature.bst}

\bibliography{bib_arxiv.bib}

\appendix
\section*{Supplementary Information}
\renewcommand{\thetable}{S\arabic{table}}
\renewcommand{\tablename}{Table}

\setcounter{table}{0}
\subsection*{Data Sources}
We report here the data sources used for the three countries, including for each dataset its description, access link, date of last visit, spatial coverage, format, and license information.

\input{data_sources_spain}

\input{data_sources_france}

\input{data_sources_italy}

\input{trade_union_data_ita}

\clearpage
\input{supplmentary_material}

\end{document}

%% file: data_sources_spain.tex
\begin{table*}[p]
\centering
\caption{Comprehensive list of input datasets used in the construction of the synthetic population for Spain, including source, access link, date of last visit, spatial resolution, and license}
\renewcommand{\arraystretch}{1.2}
\begin{tabular}{|c|p{4cm}|p{4cm}|p{1.5cm}|p{2cm}|p{1.5cm}|p{3.5cm}|}
\hline
\textbf{No.} & \textbf{Description} & \textbf{Link (Source)} & \textbf{Accessed on} & \textbf{Coverage} & \textbf{Format} & \textbf{License} \\ 
\hline

\refstepcounter{tablerow}\thetablerow\label{dat:cadastre} 
& Cadastral cartography including geolocated parcels, buildings, and administrative boundaries (coverage excludes Basque Country and Navarre). 
& \tiny\url{https://www.catastro.hacienda.gob.es/webinspire/index.html} 
& Oct 2025 
& Parcel and building level, National coverage (approx. 95\%) 
& SHP, GML 
& Publicly available (INSPIRE compliant) \\

\refstepcounter{tablerow}\thetablerow\label{dat:cadastre_navarra}
& Cadastral cartography for Navarre.
& \tiny\url{https://catastro.navarra.es/}
& Oct 2025
& Regional coverage (Navarre)
& SHP
& Publicly available \\

\refstepcounter{tablerow}\thetablerow\label{dat:b5m_gipuzkoa}
& Geographic cartography for Gipuzkoa.
& \tiny\url{https://b5m.gipuzkoa.eus/web5000/es}
& Oct 2025
& Regional coverage (Gipuzkoa)
& Multiple formats
& Publicly available \\

\refstepcounter{tablerow}\thetablerow\label{dat:cadastre_bizkaia}
& Cadastral cartography for Bizkaia.
& \tiny\url{http://arcgis.bizkaia.eus:8080/inspire/}
& Oct 2025
& Regional coverage (Bizkaia)
& WMS, WFS, ATOM
& Publicly available \\

\refstepcounter{tablerow}\thetablerow\label{dat:osm_buildings} 
& Geolocated buildings categorized by main use (residential, leisure, retail, education, offices, etc.). 
& \tiny\url{https://www.openstreetmap.org} 
& Oct 2025 
& Building level, National coverage 
& SHP 
& Open Data (ODbL, with limitations) \\

\refstepcounter{tablerow}\thetablerow\label{dat:household_size} 
& Distribution of households by number of members (1, 2, \ldots, 6+). 
& \tiny\url{https://www.ine.es/censos2011_datos/cen11_datos_resultados.htm} 
& Oct 2025 
& Municipality level, National data from 2011 census 
& CSV, XLS 
& Publicly available \\

\refstepcounter{tablerow}\thetablerow\label{dat:household_type} 
& Household composition by couple type (with and without children). 
& \tiny\url{https://www.ine.es/censos2011_datos/cen11_datos_resultados.htm} 
& Oct 2025 
& Municipality level, National data from 2011 census 
& CSV, XLS 
& Publicly available \\

\refstepcounter{tablerow}\thetablerow\label{dat:occupied_dwellings} 
& Number of occupied and unoccupied dwellings by municipality. 
& \tiny\url{https://www.ine.es/jaxi/Tabla.htm?path=/t20/e244/viviendas/p06/l0/&file=10mun00.px&L=0} 
& Oct 2025 
& Municipality level ($>$2000 inhabitants), National data from 2011 census 
& CSV 
& Publicly available \\

\refstepcounter{tablerow}\thetablerow\label{dat:population_register} 
& Population by municipality from the municipal register, disaggregated by sex. 
& \tiny\url{https://www.ine.es/jaxi/Tabla.htm?path=/t20/e244/avance/p02/l0/&file=1mun00.px&L=0} 
& Oct 2025 
& Municipality level, National coverage 
& CSV, XLS 
& Publicly available \\

\refstepcounter{tablerow}\thetablerow\label{dat:epa} 
& Economically Active Population Survey (EPA) microdata. 
& \tiny\url{https://www.ine.es/dyngs/INEbase/es/operacion.htm?c=Estadistica_C&cid=1254736176918&menu=ultiDatos&idp=1254735976595} 
& Oct 2025 
& National sample survey 
& Microdata (TXT, CSV) 
& Publicly available \\

\refstepcounter{tablerow}\thetablerow\label{dat:mobility_madrid} 
& Origin--destination trips within the Madrid Autonomous Community classified by activity type (work, study, home, shopping, etc.). 
& \tiny\url{https://datos.comunidad.madrid/catalogo/dataset/resultados-edm2018} 
& Oct 2025 
& Regional level (Madrid Autonomous Community), year 2018 
& CSV 
& Publicly available \\

\refstepcounter{tablerow}\thetablerow\label{dat:working_hours} 
& Distribution of employed population by type of working day, sex, and age group. 
& \tiny\url{https://www.ine.es/jaxiT3/Tabla.htm?t=5125&L=0} 
& Oct 2025 
& National coverage, annual data (2021--2023) 
& CSV 
& Publicly available \\

\refstepcounter{tablerow}\thetablerow\label{dat:working_time} 
& Percentage of employed persons working at each hour of the day by type of working day. 
& \tiny\url{https://www.ine.es/dynt3/inebase/index.htm?type=pcaxis&path=/t25/e447/a2009-2010/p05/&file=pcaxis&L=0} 
& Oct 2025 
& National coverage, 2009--2010 survey 
& CSV 
& Publicly available \\

\hline
\end{tabular}
\label{tab:spa_data_sources}
\end{table*}

%% file: data_sources_france.tex
\begin{table*}[p]
\centering
\caption{Comprehensive list of input datasets used in the construction of synthetic population for France, including source, access link, date of last visit, spatial resolution, and license (Part 1: 1--9).}
\label{tab:fra_data_sources}

\renewcommand{\arraystretch}{1.2}
\begin{tabular}{| p{0.5 cm}|p{7cm}|p{2cm}|p{2cm}|p{2.5cm}|p{1cm}|p{3cm}|p{3.5cm}|}
\hline
\textbf{No.} & \textbf{Description} & \textbf{Link (Source)} & \textbf{Accessed on} & \textbf{Coverage} & \textbf{Format} & \textbf{License} \\ 
\hline

1 & \textbf{Household structure}. Number of households by type (couples with/without children, single-parent families, single persons, other), including PCS and household size variables & \tiny\url{https://www.insee.fr/fr/statistiques/8268828#consulter} & March 2026 & Sub-municipality level (IRIS). & CSV & Publicly available \\

2 & \textbf{Gender split in single-parent households}. Count of single-parent families distinguishing male vs. female head of household &
\tiny\url{https://catalogue-donnees.insee.fr/fr/catalogue/recherche/DS_RP_MENAGES_COMP}
& March 2026 & Municipality level. & CSV & Publicly available \\

3 & \textbf{Age of reference person by household type}. Age class distribution of the household head conditional on household structure &
\tiny\url{https://www.insee.fr/fr/statistiques/8205233?sommaire=8205238 } 
& March 2026 & Municipality level. & CSV & Publicly available \\

4 & \textbf{Age of children and other household members}. Age distribution by role in the household (child in couple family, child in single-parent family, other member) &
\tiny\url{https://www.insee.fr/fr/statistiques/8205233?sommaire=8205238#dictionnaire} 
& March 2026 & Municipality level. & CSV & Publicly available \\

5 & \textbf{Spouse age distribution}. Microdata used to estimate the conditional distribution of the spouse's age given the reference person's age class &
\tiny\url{https://www.progedo.fr/donnees/quetelet-progedo/acceder-aux-donnees/}
& March 2026 & National level. & CSV & Not Publicly available \\

6 & \textbf{Employment status by age}. Share of active and inactive individuals by age group. &
\tiny\url{https://www.insee.fr/fr/statistiques/8268843}
\tiny\url{https://catalogue-donnees.insee.fr/fr/catalogue/recherche/DS_RP_EMPLOI_LR_PRINC}
& March 2026 & Sub-municipality level (IRIS). & CSV & Publicly available \\

7 & \textbf{Student status by age}. Share of students by age group. &
\tiny\url{https://catalogue-donnees.insee.fr/fr/catalogue/recherche/DS_RP_EMPLOI_LR_PRINC}
& March 2026 & Municipality level. & CSV & Publicly available \\

8 & \textbf{Household size by age}. Distribution of household size (number of persons) by age class of the reference person. &
\tiny\url{https://www.insee.fr/fr/statistiques/8205233?sommaire=8205238#consulter}
& March 2026 & Municipality level. & CSV & Publicly available \\

9 & \textbf{Building stock}. Geometry, number of floors, number of dwellings and floor area for all buildings; used to assign households to physical locations. &
\tiny\url{https://bdnb.io/archives_data/bdnb_millesime_2025_07_a/} / 
\tiny\url{https://geoservices.ign.fr/contoursiris}
& March 2026 & Building. & CSV & Publicly available \\

\hline
\end{tabular}
\end{table*}

\begin{table*}[p]
\centering
\caption{Comprehensive list of input datasets used in the construction of synthetic populations, including source, access link, date of last visit, spatial resolution, and license. (Part 2: 10--18)}
\label{tab:fra_data_sources_2}

\renewcommand{\arraystretch}{1.2}
\begin{tabular}{| p{0.5 cm}|p{7cm}|p{2cm}|p{2cm}|p{2.5cm}|p{1cm}|p{3cm}|p{3.5cm}|}
\hline
\textbf{No.} & \textbf{Description} & \textbf{Link (Source)} & \textbf{Accessed on} & \textbf{Coverage} & \textbf{Format} & \textbf{License} \\ 
\hline

10 & \textbf{Business establishments (SIRET)}. Establishment location (Lambert-93 coordinates) and workforce size bracket, used to generate workplace locations. &
\tiny\url{https://www.data.gouv.fr/datasets/base-sirene-des-entreprises-et-de-leurs-etablissements-siren-siret} / 
& March 2026 & Establishment level & CSV & Publicly available \\

11 & \textbf{School and community locations}. Buildings identified as schools (nursery, primary, middle, high school, university) and community (EHPAD, social residences, university residences) extracted from the BDNB equipment layer &
\tiny\url{https://bdnb.io/archives_data/bdnb_millesime_2025_07_a/} / 
& March 2026 & Establishment level & CSV & Publicly available \\

12 & \textbf{Work commuting flows}. Origin–destination matrix of active workers between municipalities &
\tiny\url{https://www.insee.fr/fr/statistiques/8201899} / 
& March 2026 & Municipality & CSV & Publicly available \\

13 & \textbf{School commuting flows}. Origin–destination matrix of students by school level (nursery to university) &
\tiny\url{https://www.insee.fr/fr/statistiques/8205892?sommaire=8205966&q=mobilit%C3%A9%20scolaire#consulter-sommaire} / 
& March 2026 & Municipality & CSV & Publicly available \\

14 & \textbf{IRIS population structure microdata}. Number of people by sex and age class in each IRIS &
\tiny\url{https://www.insee.fr/fr/statistiques/8202264?sommaire=8202287} / 
& June 2026 & Municipality & CSV & Publicly available \\

15 & \textbf{Age-sex pyramid}. National number of people by age and sex &
\tiny\url{https://www.insee.fr/fr/statistiques/6038698} / 
& June 2026 & National & CSV & Publicly available \\

16 & \textbf{IRIS geographical limits}. Shape of IRIS &
\tiny\url{https://cartes.gouv.fr/rechercher-une-donnee/dataset/IGNF_CONTOURS-IRIS?redirected_from=geoservices.ign.fr} / 
& June 2026 & IRIS & GPKG & Publicly available \\

17 & \textbf{Population living in collective houses}. Number of people living in collective housing &
\tiny\url{https://www.insee.fr/fr/statistiques/8205892} / 
& June 2026 & IRIS & GPKG & Publicly available \\

18 & \textbf{Working schedules data}. Part of the working population working partial time, on Saturday, Sunday and at night &
\tiny\url{https://dares.travailemploi.gouv.fr/donnees/portraits-statistiques-des-metiers} / 
& September 2026 & National & CSV & Publicly available \\

\hline

\end{tabular}
\end{table*}

%% file: data_sources_italy.tex
\begin{table*}[p]
\centering
\caption{Comprehensive list of input datasets used in the construction of synthetic populations for Italy, including source, access link, date of last visit, spatial resolution, and license.}
\label{tab:ita_data_sources}
\renewcommand{\arraystretch}{1.2}
\begin{tabular}{|c|p{4cm}|p{4cm}|p{1.5cm}|p{2cm}|p{1.5cm}|p{3.5cm}|}
\hline
\textbf{No.} & \textbf{Description} & \textbf{Link (Source)} & \textbf{Accessed on} & \textbf{Coverage} & \textbf{Format} & \textbf{License} \\ 
\hline
\refstepcounter{tablerow}\thetablerow\label{dat:microdata} & Family composition microdata. & \tiny\url{https://www.istat.it/microdati/censimento-popolazione-e-abitazioni-2001/} & Sep 2026 & Municipality to National (2\% sample) from 2001 census & CSV & Restricted Access (Available on Request) \\
\refstepcounter{tablerow}\thetablerow\label{dat:census_tract} & Number of residents by sex and age, and dwellings by size. & \tiny\url{https://www.istat.it/notizia/dati-per-sezioni-di-censimento/} & Sep 2026 & Census tract to whole Nation from 2021 census & XLSX & Publicly available \\
\refstepcounter{tablerow}\thetablerow\label{dat:boundaries} & Census tract to Macroregional Level Boundaries.  & \tiny\url{https://www.istat.it/notizia/confini-delle-unita-amministrative-a-fini-statistici-al-1-gennaio-2018-2/} & Sep 2026 & Census tract to whole Nation from 1991 to 2025 census. & SHP & Publicly available \\
\refstepcounter{tablerow}\thetablerow\label{dat:localunits} & Number of units and workers per census tract from 2011 census. & \tiny\url{https://www.istat.it/storage/cartografia/variabili-censuarie/dati-sce_2011.zip} Landing page: \tiny\url{https://www.istat.it/notizia/basi-territoriali-e-variabili-censuarie/} & Sep 2026 & Census Tract, National Data & TXT & Publicly available \\
\refstepcounter{tablerow}\thetablerow\label{dat:working_stats} & Distribution of working-sector by gender and municipality. & \tiny\url{https://esploradati.istat.it/databrowser/#/it/censpop/categories/IT1,Z1200CPA,1.0/APPR/EMPLP/IT1,DF_DCSS_EMPLP_2_COM,1.0} & Sep 2026 & Municipality, National Data as on 31st December 2021. & CSV, XLSX, JSON, SDMX & Publicly available \\
\refstepcounter{tablerow}\thetablerow\label{dat:professional_condition} & List of workers, students and unemployed by age and sex and municipality as on 31st December 2022. & \tiny\url{https://esploradati.istat.it/databrowser/#/it/censpop/categories/IT1,Z1200CPA,1.0/CPA_POP/DCSS_ISTR_LAV_PEN_2_TV/IT1,DF_DCSS_ISTR_LAV_PEN_2_TV_3,1.0} & Sep 2026 & Municipality level, National data & CSV, XLSX, JSON, SDMX & Publicly available \\
\refstepcounter{tablerow}\thetablerow\label{dat:schools} & List of schools by degree with addresses. & \tiny\url{https://dati.istruzione.it/opendata/opendata/catalogo/elements1/?area=Scuole} & Sep 2026 & Municipality level, National data from 2024 school census. & CSV, JSON, RDF, XML & Publicly available \\ 
\refstepcounter{tablerow}\thetablerow\label{dat:students} & List of schools by degree with students, including infancy enrollment from a separate dataset in the same catalog (partial coverage). & \tiny\url{https://dati.istruzione.it/opendata/opendata/catalogo/elements1/?area=Scuole} & Sep 2026 & Municipality level, National data from 2024 school census. & CSV, JSON, RDF, XML & Publicly available \\
\refstepcounter{tablerow}\thetablerow\label{dat:od} & Origin Destination Matrix by professional condition (Students, Workers). & \tiny\url{https://www.istat.it/storage/cartografia/matrici_pendolarismo/matrici_pendolarismo_2011.zip} Landing page: \url{https://www.istat.it/non-categorizzato/matrici-del-pendolarismo/} & Sep 2026 & Census Tract, National data from 2011 survey. & TXT & Publicly available \\ 
\refstepcounter{tablerow}\thetablerow\label{dat:pyramid} & Age and sex pyramid by municipality. & \tiny\url{https://demo.istat.it/app/?l=it&a=2021&i=POS} & Sep 2026 & Municipality, National data from 2021 demographic records. & CSV & Publicly available \\ 

\hline

\hline
\end{tabular}
\end{table*}

%% file: trade_union_data_ita.tex
\begin{table*}[p]
\centering
\caption{Working schedules by ATECO macrosector used in the construction of synthetic populations for Italy, including weekday, Saturday, and Sunday schedules, and source (Part 1: sectors A--K).}
\label{tab:ita_working_schedules_1}
\renewcommand{\arraystretch}{1.2}
\begin{tabular}{|p{3cm}|c|p{2.5cm}|p{2.5cm}|p{2.5cm}|p{2.5cm}|p{3cm}|}
\hline
\textbf{Macrosector} & \textbf{ATECO} & \textbf{Notes} & \textbf{Weekday Schedule} & \textbf{Saturday Schedule} & \textbf{Sunday Schedule} & \textbf{Source} \\
\hline
Agriculture, Forestry, Fish Farming & A & 39 hours/week, 6.5 hours/day & 05:30--12:00, 06:00--12:30, 06:30--13:00 & 05:30--12:00, 06:00--12:30, 06:30--13:00 & None & \tiny\url{https://www.unimpresa.it/ccnl/ccnl-lavoratori-agricoli} \\
\hline
Extraction Activities & B & 40 hours/week, 8 hours/day & 06:00--14:00, 14:00--22:00, 22:00--06:00 & 06:00--14:00, 14:00--22:00, 22:00--06:00 & 06:00--14:00, 14:00--22:00, 22:00--06:00 & \tiny\url{https://www.studiocerbone.com/ccnl-aziende-esercenti-le-attivita-escavazione-lavorazione-dei-materiali-lapidei/} \\
\hline
Manufacturing Activities & C & 40 hours/week, 8 hours/day & 06:00--14:00, 14:00--22:00, 22:00--06:00 & 06:00--14:00, 14:00--22:00, 22:00--06:00 & 06:00--14:00, 14:00--22:00, 22:00--06:00 & \tiny\url{https://www.dipendentincloud.it/guida-gestione-turni-lavoro/lavoro-turni/} \\
\hline
Supply of Electricity, Gas, Steam, and Air Conditioning & D & 40 hours/week, 8 hours/day & 06:00--14:00, 14:00--22:00, 22:00--06:00 & 06:00--14:00, 14:00--22:00, 22:00--06:00 & 06:00--14:00, 14:00--22:00, 22:00--06:00 & \tiny\url{https://www.dipendentincloud.it/guida-gestione-turni-lavoro/lavoro-turni/} \\
\hline
Supply of Water; Sewage, Waste Treatment and Sanitation & E & 40 hours/week, 8 hours/day & 06:00--14:00, 14:00--22:00, 22:00--06:00 & 06:00--14:00, 14:00--22:00, 22:00--06:00 & 06:00--14:00, 14:00--22:00, 22:00--06:00 & \tiny\url{https://www.dipendentincloud.it/guida-gestione-turni-lavoro/lavoro-turni/} \\
\hline
Construction & F & & 08:00--13:00, 15:00--19:00 & None & None & \tiny\url{https://www.infobuild.it/opere-edili-in-condominio-gli-orari-da-rispettare/} \\
\hline
Trade & G & 40 hours/week, 5 days & 09:00--15:00, 15:00--21:00 & 09:00--15:00, 15:00--21:00 & 09:00--15:00, 15:00--21:00 & \tiny\url{https://www.confcommercio.it/-/ccnl-terziario-distribuzione-servizi-testo-unico-2019} \\
\hline
Transportation and Stocking & H & 40 hours/week, 5 days & 06:00--14:00, 14:00--22:00, 22:00--06:00 & 06:00--14:00, 14:00--22:00, 22:00--06:00 & 06:00--14:00, 14:00--22:00, 22:00--06:00 & \tiny\url{https://www.world-matic.com/blog/turni-amazon-massacranti/} \\
\hline
Accommodation and Catering Services & I & & 07:00--15:00, 15:00--23:00, 23:00--07:00 & 07:00--15:00, 15:00--23:00, 23:00--07:00 & 07:00--15:00, 15:00--23:00, 23:00--07:00 & \tiny\url{https://hoteltechreport.com/it/news/hotel-front-desk} \\
\hline
Information and Communication Services & J & 36 hours/week, 5 days & 09:00--17:00 & None & None & \tiny\url{https://www.odg.it/wp-content/uploads/2016/08/Testo-del-Contratto.pdf} \\
\hline
Financial Services & K & 37.5 hours/week, 5 days & 09:00--18:00 & None & None & \tiny\url{https://www.ivass.it/chi-siamo/in-sintesi/lavorare-in-ivass/carriera-e-rapporto-di-impiego/index.html} \\
\hline
\end{tabular}
\end{table*}

\begin{table*}[p]
\centering
\caption{Working schedules by ATECO macrosector used in the construction of synthetic populations for Italy, including weekday, Saturday, and Sunday schedules, and source (Part 2: sectors L--U).}
\label{tab:ita_working_schedules_2}
\renewcommand{\arraystretch}{1.2}
\begin{tabular}{|p{3cm}|c|p{2.5cm}|p{2.5cm}|p{2.5cm}|p{2.5cm}|p{3cm}|}
\hline
\textbf{Macrosector} & \textbf{ATECO} & \textbf{Notes} & \textbf{Weekday Schedule} & \textbf{Saturday Schedule} & \textbf{Sunday Schedule} & \textbf{Source} \\
\hline
Real Estate Services & L & & 09:00--20:00 & None & None & \tiny\url{https://tempocasa.it/blog/agente-immobiliare-nel-vivo-della-professione-tra-orari-ritmi-appuntamenti-compensi-e-provvigioni/} \\
\hline
Professional, Scientific and Technical Services & M & & 09:00--18:00 & None & None & --- \\
\hline
Rental, Travel Agencies and Business Support Services & N & 40 hours/week & 09:00--18:00 & None & None & \tiny\url{https://www.filcams.cgil.it/page/agenzie_viaggi} \\
\hline
Public Administration and Defence; Compulsory Social Security & O & 48 hours/week & 09:00--19:00 & None & None & \tiny\url{https://www.difesa.it/sgd-dna/staff/dg/persociv/faq/orario-lavoro/33360.html} \\
\hline
Education & P & & 08:00--16:00 & None & None & \tiny\url{https://osservatoriocpi.unicatt.it/ocpi-pubblicazioni-le-ore-di-lavoro-dei-docenti-italiani-i-risultati-di-una-nostra-indagine} \\
\hline
Healthcare & Q & & 06:00--14:00, 14:00--22:00, 22:00--06:00 & 06:00--14:00, 14:00--22:00, 22:00--06:00 & 06:00--14:00, 14:00--22:00, 22:00--06:00 & --- \\
\hline
Artistic, Sporting, and Entertainment Activities & R & & 06:00--14:00, 14:00--22:00, 22:00--06:00 & 06:00--14:00, 14:00--22:00, 22:00--06:00 & 06:00--14:00, 14:00--22:00, 22:00--06:00 & \tiny\url{https://www.confcommercio.it/-/ccnl-terziario-distribuzione-servizi-testo-unico-2019} \\
\hline
Other Services & S & & 06:00--14:00, 14:00--22:00, 22:00--06:00 & 06:00--14:00, 14:00--22:00, 22:00--06:00 & 06:00--14:00, 14:00--22:00, 22:00--06:00 & \tiny\url{https://www.confcommercio.it/-/ccnl-terziario-distribuzione-servizi-testo-unico-2019} \\
\hline
Activities of Families and Households as Employers of Domestic Staff & T & 40 hours/week & 06:00--14:00, 14:00--22:00 & None & None & \tiny\url{https://www.safacli.com/guida-al-contratto-di-lavoro-domestico/} \\
\hline
Extraterritorial Organizations and Bodies & U & No agents assigned to this activity & N/A & N/A & N/A & N/A \\
\hline
\end{tabular}
\end{table*}

%% file: supplmentary_material.tex
\section{Data Sources}
\input{data_source}

\section{Companion datasets}

\subsection{Buildings}

The locations of the activities performed by the agents are provided in the buildings dataset, described in \autoref{tab:buildings_fields}. Each row corresponds to a place that an agent can be assigned to: a dwelling, a workplace, a school or another point of interest. Each place is identified by the \texttt{reference} field, which links it to the agents' homes (\texttt{dwelling\_id}) and main activity destinations (\texttt{ref\_dest}) in \autoref{tab:output_fields}. Every place is described at multiple spatial scales: the municipality (\texttt{code}) and the census tract or cell (\texttt{cell\_group}) containing it, and the projected coordinates of its centroid (\texttt{coordinate}). Places are classified by sector (residential, generic commercial/industrial, or point of interest), and points of interest are further specified by a category (\texttt{LayerIndex}, see \autoref{tab:sector_definitions}) and a free-text subcategory. The \texttt{value} field gives a country-specific measure of the size or capacity of the place, which reflects the different data sources available in each country: real building footprints from cadastral and OpenStreetMap data for Spain, building footprints and number of floors from the national building database for France, and synthetic placements within census tracts, sized from local-unit data, for Italy. As the coordinate reference system also differs by country, the coordinates should be reprojected to a common system (e.g.\ WGS~84) before combining data from different countries.
 
\begin{table*}[ht]
\centering
\caption{%
  Buildings dataset. This table contains the description of the buildings and points of interest location at multiple scales, sector and category. The \texttt{reference} field identifies the same objects defined in the fields \texttt{dwelling\_id} and \texttt{ref\_dest} in \autoref{tab:output_fields}.
}
\label{tab:buildings_fields}
\small
\begin{tabular}{p{3.2cm} p{1.8cm} p{1.4cm} p{6.8cm}}
\hline
\textbf{Field} & \textbf{Type} & \textbf{Nullable} & \textbf{Description} \\
\hline
\texttt{reference}
  & string & No & Unique building identifier; matching \texttt{dwelling\_id} and \texttt{ref\_dest} in the agents' dataset in \autoref{tab:output_fields}. \\[4pt]
\texttt{code}
  & string & No & Links to the \texttt{code} field in \autoref{tab:output_fields}.. \\[4pt]
\texttt{cell\_group}
  & string & No & Census-tract/cell identifier. \\[4pt]
\texttt{sector}
  & categorical & No & Building-level sector: 1~residential, 2~commercial/industrial (generic), 3~point of interest. Links to the \texttt{sector} field in \autoref{tab:output_fields} \\[4pt]
\texttt{LayerIndex}
  & categorical & Yes & Point-of-interest category, populated only when \texttt{sector}~=~3; see \autoref{tab:sector_definitions}. \\[4pt]
\texttt{Subcategory}
  & string & Yes & Free-text label further specifying \texttt{LayerIndex} (e.g.\ \texttt{'restaurant'}, \texttt{'hospital'}). \\[4pt]
\texttt{value}
  & integer & No & For Spain, building footprint area, in m\textsuperscript{2}; for Italy, number of workers and students estimated from local units data; for France, the product between building footprint area and the number of floors.\\[4pt]
\texttt{coordinate}
    & (float, float) & No & Easting and northing of the building centroid, in metres, in a projected coordinate reference system that differs by country: for Spain, Madrid 1870 (Madrid) / Spain LCC (EPSG:2062), except for the Canary Islands (provinces 35 and 38), which use REGCAN95 / LAEA Europe (EPSG:5635); for France, RGF93 / Lambert-93 (EPSG:2154); for Italy, WGS~84 / UTM zone 32N (EPSG:32632). \\[4pt]
\hline
\end{tabular}
\end{table*}

The classification of places adopted in the buildings dataset is summarised in \autoref{tab:sector_definitions}. At the coarsest level, the \texttt{sector} field, shared with the agents' dataset (\autoref{tab:output_fields}), distinguishes residential buildings (sector~1), generic agricultural, commercial and industrial workplaces (sector~2) which are accessible only to the employed agents, and points of interest (sector~3). The latter are places that, besides employing workers, are visited by other agents, for example to attend school, receive health care, shop or spend leisure time; for this reason they are further classified through the \texttt{LayerIndex} field into a common set of categories, such as education and public services, healthcare, leisure, food and drink, accommodation and retail. The \texttt{LayerIndex} categories were defined to be shared across countries, so that points of interest derived from different national sources (business registers, local-unit data and OpenStreetMap) can be compared and used consistently in epidemic models; the more detailed, source-specific label of each point of interest is kept in the \texttt{Subcategory} field. Since the available sources differ by country, not every category is populated in every national dataset.

\begin{table*}[ht]
\centering
\caption{%
  National definitions of the \texttt{sector} (\autoref{tab:output_fields}, \autoref{tab:buildings_fields}) and \texttt{LayerIndex} (\autoref{tab:buildings_fields}) categories.
}
\label{tab:sector_definitions}
\small
\begin{tabular}{p{3.2cm} p{3.2cm} p{6.8cm}}
\hline
\textbf{sector} & \textbf{LayerIndex} & \textbf{Label} \\
\hline
1 & --- & Residential \\[2pt]
2 & --- & Agriculture / commercial / industrial (generic) \\[2pt]
3 & 0 & Education / public services \\[2pt]
3 & 1 & Healthcare \\[2pt]
3 & 2 & Leisure \\[2pt]
3 & 3 & Food \& drink \\[2pt]
3 & 4 & Accommodation \\[2pt]
3 & 5 & Retail \\[2pt]
3 & 6 & Finance \\[2pt]
3 & 7 & Tourism / heritage \\[2pt]
3 & 9 & Other \\[2pt]
3 & 10 & Religious \\
\hline
\end{tabular}
\par\vspace{4pt}
\end{table*}

\subsection{Schedules}

For each country, the synthetic population is complemented by an activity schedules dataset, described in \autoref{tab:schedules_fields}. The dataset contains one row per agent; the $i$-th row of the schedules file therefore refers to the $i$-th agent of the corresponding population file, while the home dwelling is also reported explicitly through the \texttt{home\_id} field.
For each agent, the dataset reports the daily time window of the main activity (work or school), including a second interval for split shifts, the days of the week on which this activity takes place, and the type of shift.
It also provides the home and destination locations, both as projected coordinates and as latitude/longitude, the home-to-destination distance, and the census tracts of residence and of the main activity.
Finally, to support the modelling of non-routine mobility, each agent is linked to the nearest hospital and to the three nearest supermarkets, leisure venues and restaurants or bars around their home.
Fields related to the main activity (shift times, destination coordinates, destination type and distance) are null or zero for agents without an occupation. 
The schema is shared among Italy, Spain and France, with a few country-specific differences: the nearest points of interest (\texttt{ref\_hosp}, \texttt{sup\_*}, \texttt{lei\_*}, \texttt{rest\_*} and the corresponding counters) and the destination type \texttt{ins\_type} are available for Italy and Spain only, whereas for France the destination is characterised by its economic activity code (\texttt{ATECO}); the transport mode (\texttt{means\_trans}) is populated for Spain only.

\begin{table*}[ht]
\centering
\caption{%
  Activity schedules dataset.
  The dataset contains one row per active agent's daily activity, describing their time schedules, and the closest points of interest from agent's home. Unless otherwise noted, fields marked Nullable~=~Yes are null for unoccupied agents.
}
\label{tab:schedules_fields}
\small
\begin{tabular}{p{3.2cm} p{1.8cm} p{1.4cm} p{6.8cm}}
\hline
\textbf{Field} & \textbf{Type} & \textbf{Nullable} & \textbf{Description} \\
\hline
\texttt{dwelling\_id}
  & string & No & Links to the \texttt{dwelling\_id} field in \autoref{tab:output_fields}. \\[4pt]
\texttt{person\_id}
  & integer & No & Links to the \texttt{person\_id} field in \autoref{tab:output_fields}. \\[4pt]

\texttt{sector}
  & categorical & Yes & Links to the \texttt{sector} field in \autoref{tab:output_fields}. \\[4pt]
\texttt{tt\_hini1}, \texttt{tt\_hini2}, \texttt{tt\_hend1}, \texttt{tt\_hend2}
  & float & Yes & Start/end hour(s) of the agent's daily work/school shift; the \texttt{\_2} fields are populated only for split-shift schedules. \\[4pt]
\texttt{tt\_monday} \ldots \texttt{tt\_sunday}
  & categorical & No & 0/1 flag for each day of the week, indicating whether the shift applies that day. \\[4pt]
\texttt{tt\_type}
  & categorical & Yes & Shift-type code. \\[4pt]
\texttt{x}, \texttt{y}
  & float & No & Home building coordinates, in the national projected coordinate reference system (matching the buildings dataframe's \texttt{geometry} field). \\[4pt]
\texttt{h\_lat}, \texttt{h\_lon}
  & float & No & Home building coordinates, latitude/longitude. \\[4pt]
\texttt{home\_id}
  & string & No & Home dwelling identifier; equals \texttt{dwelling\_id} in the agents dataframe. \\[4pt]
\texttt{ref\_dest}
  & string & No & Workplace/school identifier for active agents, own \texttt{home\_id} for unoccupied agents; equals \texttt{ref\_dest} in the agents dataframe. \\[4pt]
\texttt{ins\_lat}, \texttt{ins\_lon}
  & float & Yes & Destination (workplace/school) building coordinates, latitude/longitude. \\[4pt]
\texttt{cell\_group}
  & string & No & Census-tract/cell of residence; equals \texttt{cell\_group} in the agents dataframe. \\[4pt]
\texttt{cell\_dest}
  & string & No & Census-tract/cell of the main activity location; equals \texttt{cell\_dest} in the agents dataframe. \\[4pt]
\texttt{ins\_type}
  & categorical & Yes & Numeric code of the destination building's type, matching the taxonomy of the buildings dataframe. \\[4pt]
\texttt{distance\_hw}
  & float & Yes & Home-to-workplace/school distance. \\[4pt]
\texttt{is\_student}
  & categorical & No & Flag indicating whether the agent is a student. \\[4pt]
\hline
\end{tabular}
\end{table*}

\section*{Synthetic population generation}~\label{appendix:methods}

\subsection*{Spain}

A set of statistical distributions was generated from the input data described in the previous section. These distributions provide the basis for assigning socio-demographic and economic attributes to the synthetic population of agents. Specifically, the following distributions were derived:

\begin{itemize}

\item \textbf {Household type}: obtained from household microdata, disaggregated by province code and municipality size.

 \item \textbf{Household size:} derived from household microdata, conditioned on province code, municipality size, and household type.

\item \textbf{Age:} based on individual-level microdata, considering province code, municipality size, household type, and individual identifiers.

\item \textbf{Sex:} estimated from individual-level microdata, disaggregated by province code, municipality size, household type, and person identifier.

\item \textbf{Activity status:} derived from individual-level microdata, based on province code, municipality size, household type, and age.

\item \textbf{Working sector:} obtained from labor force survey (EPA) microdata, by province code and sex.

\item \textbf{Regional working sectors: }refined using building floor-area data, allowing the distribution of workers across economic sectors according to the built environment in each region.

\item \textbf{Mobility patterns:} constructed from mobility datasets, providing probabilities of travel between spatial cells.

\item \textbf{Workplace assignment:} estimated as the probability of working in a given building, considering its location (statistical cell), sector, and area.

\end{itemize}

Land-use information was further enriched by combining cadastral data with OpenStreetMap (OSM), allowing the classification of parcels into categories such as residential, workplaces (by type), retail, leisure services (e.g., restaurants, cinemas), hospitals, and schools. Once the parcels were characterized, households were first allocated to residential buildings. After spatially distributing the population, the remaining attributes of the individuals (homeowner type, household size, age, sex, activity status) were assigned according to the previously derived distributions.

Origin–destination regions were then defined using the spatial grid division provided by the Spanish National Statistics Institute (INE). Each agent was associated with an origin region corresponding to the grid cell where their residence is located (previously assigned to a specific cadastral dwelling) and with a destination region representing their workplace area. Destination regions were determined using experimental origin–destination mobility flows derived from anonymized mobile phone data provided by INE.

\subsubsection*{Activity assignation}

Once destination regions were assigned, employment sectors were allocated to agents. This step combines the statistical distributions of the number of workers per sector with the spatial distribution of buildings associated with each sector identified in the cadastral data within each grid cell. Assigning sectors after determining the destination region allows the model to simultaneously respect the overall distributions of workers by sector while ensuring that workers are allocated proportionally to the number of sector-specific buildings available in each cell. This prevents the assignment of disproportionate numbers of workers in sectors for which there is insufficient corresponding infrastructure in the destination area.

\subsubsection*{Schedules}

Working schedules were assigned to agents using the distributions derived from the Spanish Time Use Survey (Encuesta de Empleo del Tiempo) conducted by INE. This allows the synthetic population to reproduce realistic daily activity patterns and commuting dynamics.

The overall process follows a consistent methodology across countries, including France and Italy, ensuring comparability of the generated populations.

\subsection*{France}

\subsubsection*{Data preprocessing}

We describe here the preprocessing part of the pipeline, to produce probability distributions from raw data referred to~\autoref{tab:fra_data_sources}.
\\

\paragraph*{Household Type Distribution.}

Seven mutually exclusive household types are defined: single male (HH=1), single female (HH=2), couple without children (HH=3), couple with children (HH=4),  single father (HH=5), single mother (HH=6), other (HH=7). For each IRIS, the frequency of each type is computed from \textit{Data Source 1} , which directly provides counts for each categories.

However, single-parent households are only recorded as a combined total without the gender of the parent. To recover the male/female split, a municipality-level gender ratio is derived from \textit{Data Source 2}, using household types MF221 (male head) and MF222 (female head). This ratio is then applied at IRIS level under the assumption that the gender split within an IRIS mirrors that of its parent municipality.

\textbf{Missing data:} Some municipalities are absent from \textit{Data Source 2}. Two cases are distinguished: (A) if the IRIS has no single-parent households , we assume that P(HH=6) = P(HH=7) = 0; (B) otherwise, the ratio is imputed with the mean ratio computed across all municipalities with non-missing, non-zero values.
\\

\paragraph*{Age of the Reference Person by Household Type.}
The file \textit{Data Source 3} provides, at municipality level, the count of households by household type and age class of the reference person (7 classes: <20, 20–24, 25–39, 40–54, 55–64, 65–79, 80+). Since \textit{Data Source 3} only provides the age distribution for the aggregated single-person (code 100) and single-parent (code 400) categories, \textit{Data Source 4} is used to disaggregate these by gender.

\textbf{Missing data:} Missing counts in both files are imputed by the group mean over all municipalities sharing the same household type and age class. Municipalities missing entirely from the dataset are filled with the overall mean distribution. All frequencies are computed at municipality level and then broadcast to IRIS level under the assumption that the age distribution is homogeneous within a municipality.
\\

\paragraph*{Age of Children and Other Household Members.}

From \textit{Data Source 4}, the age distribution of non-reference-person household members is derived for three roles: child in a couple family, child in a single-parent family, and other member. Only male records are used, under the assumption that the age distribution does not differ by gender for these roles. Frequencies are computed at municipality level and broadcast to IRIS.
\\

\paragraph*{Spouse Age Distribution.}

Using \textit{Data Source 5}, the joint distribution of the reference person's age class and the spouse's age class is estimated nationally. The conditional distribution P(partner age | reference age) is computed and replicated identically for all IRIS, as no sub-national data is available for this variable.
\\

\paragraph*{Employment Status by Age.}

Three activity statuses are considered: active, student, and inactive. For each IRIS, the probability of being active is computed directly from \textit{Data Source 6} for three age groups (15–24, 25–54, 55–64). Individuals between 3 and  15 are assumed to be students; those 65 and over and those under 3 are assumed inactive. The probability of being a student conditional on being inactive is estimated at municipality level from \textit{Data Source 7}, using activity code 33 (student) as a share of all non-employed individuals, separately for age groups 15–24, 25–54, and 55–64. The probability of being inactive is then derived as the complement: 

$P(\text{inactive}) = (1 - P(\text{active})) \times (1 - P(\text{student} \mid \text{not active}))$

\textbf{Missing data:} Municipality-level student rates missing from \textit{Data Source 7}, are imputed by the mean over all municipalities for the corresponding age group.
\\

\paragraph*{Household Size by Age of the reference person.}

From \textit{Data Source 8}, the distribution of household size conditional on the age class of the reference person is estimated at municipality level.

\paragraph*{Building Stock Preprocessing}

Building data from \textit{Data Source 9} is loaded department by department. Each building is assigned to an IRIS via a two-step spatial join against the IGN IRIS boundary file (Lambert-93 projection): first using the representative point of the building polygon (more robust than the centroid for concave shapes), then via maximum-intersection area for buildings that remain unmatched. Buildings with no IRIS assignment are dropped.
For IRIS where the total number of dwellings across buildings falls below the census household count (from \textit{Data Source 1}), the deficit is distributed randomly across buildings using a multinomial draw with equal probabilities. The habitable surface per dwelling is estimated as $\mathrm{floor \ area} \times \mathrm{number \ of \ floors} / \mathrm{number \ of \ dwellings}$.
\\

\paragraph*{Workplace and School Location.}

Workplace locations are derived from the SIRET database (\textit{Data Source 10}). Establishments with missing coordinates or zero-bracket workforce are excluded. Each establishment is matched to its nearest building using a k-d tree on Lambert-93 coordinates. A random workforce size is drawn uniformly within the bracket bounds, and the probability of working at a given establishment is set proportional to its workforce share within the municipality.
School locations are extracted from the BDNB equipment layer (\textit{Data Source 11}) by matching building equipment type labels  against reference lists for each school level (nursery, primary, middle school, high school, university). The probability of attending a particular establishment is set proportional to its total floor area within the municipality and school type.
\\

\paragraph*{Commuting Flow Matrices.}

For work commuting, origin–destination flows between municipalities are taken from the 2021 census mobility file (\textit{Data Source 12}). Only destinations with at least one registered workplace are retained (from previous step). For municipalities absent from the flow file, the probability distribution over destinations is imputed using the aggregate flow distribution of the corresponding department. A similar procedure is applied for school commuting, separately for each school level, using the 2021 school mobility microdata file \textit{Data Source 13}.
\\

\paragraph*{Population Pool.} Beyond the per-IRIS probability distributions described above, the pipeline requires an explicit pool of individuals to be drawn from during generation. From \textit{Data Source 14}, the resident population of each IRIS is extracted by sex and by age class (10 classes: 0–2, 3–5, 6–10, 11–17, 18–24, 25–39, 40–54, 55–64, 65–79, 80+). Because the source reports population counts by age class but the generation requires single-year ages, each class is expanded to exact ages using the national age pyramid (\textit{Data Source 15}) as the within-class age profile: for each (age class, sex), the national distribution of single-year ages within the class bounds provides weights that split the IRIS class count into per-year counts. To guarantee that class totals are preserved exactly after rounding to integers, a largest-remainder method is used (per-year counts are floored, and the residual units are allocated to the years with the largest fractional parts). The resulting pool, indexed by (IRIS, exact age, sex), is the population reservoir consumed by the generation stage: reference persons, spouses, and children draw their exact age and sex from it, and each placement decrements the corresponding entry.
\\

\paragraph*{Collective Housing.} Individuals living in communal establishments (student residences, retirement homes, care facilities, workers' hostels, and other communities) are handled separately from ordinary households. Two ingredients are prepared. First, the number of people in collective housing per department, age class, and sex is computed from the individual-level census file (\textit{Data Source 17}), by counting records whose cohabitation mode indicates communal living, aggregated to the (department, age class, sex) level. Second, the physical establishments are extracted from the BDNB equipment layer (\textit{Data Source 11}): building equipment labels are matched against reference lists to classify each collective building into a type (retirement home, student residence, disability facility, social facility, medical-care facility, or other community), and each establishment is georeferenced and attached to its commune and IRIS through the same spatial procedure used for the building stock. At the generation stage, the target counts are used to draw the corresponding number of individuals from the residual pool (those not placed in ordinary households), and each drawn individual is assigned to the nearest establishment of the type appropriate to their age class.

\subsubsection*{Algorithm of synthetic population generation}

The synthetic population is generated at the department level through a sequential probabilistic procedure, drawing on probability distributions computed from data, as explained in the previous section. The algorithm proceeds in two stages: household generation and individual assignment.
\\

\paragraph*{Household generation.} For each IRIS, we generate the same number of households as in the census data. Their attributes are drawn sequentially, each conditioned on the previous: household type,  reference-person age class, spouse age class (for couples),  household size, socio-professional category, and finally income. Every categorical draw uses a stratified multinomial allocation, so realized marginals match the target probabilities up to rounding rather than fluctuating independently across households.
\\

\paragraph*{Exact age and sex.} The exact age and sex of the reference person and spouse are obtained by drawing an actual individual from the IRIS population pool: within the age class, an exact age is sampled proportionally to the residual local counts, and the individual is removed from the pool. Households of an IRIS thus share a finite reservoir. When a class is locally exhausted, the draw relaxes to adjacent classes before being abandoned; a household whose reference person cannot be drawn is discarded. Since relaxation may shift the age outside the initial class, age classes are recomputed from the realised ages.
\\

\paragraph*{Individual assignment.} Individuals are assembled from the household attributes. Additional members (children and others) receive an exact age and sex from the same pool, subject to a constraint: in couple-with-children and single-parent households a member must be at least fifteen years younger than the youngest parent, while "other" households are unconstrained. The most constrained members are drawn first to limit placement failures; a member with no admissible individual left in the pool is dropped. Activity status is then assigned from the age-conditional employment distribution, with young children forced to inactive.
\\

\paragraph*{Collective housing.} Per-department target counts by age class and sex determine how many individuals are drawn at random from the residual pool into communal establishments. Each is assigned to the nearest establishment of the type matching their age class—retirement homes for the oldest classes, student residences for young adults, generic facilities otherwise, preferring their own commune. Individuals with no suitable establishment in the department stay in the pool and are excluded. Each forms a single-person household of a dedicated type (household type = 9).
\\

\paragraph*{Dwelling and destination.} Ordinary households are matched to dwellings within their IRIS by rank—larger and, at equal size, wealthier households to larger dwellings. Each individual is then given a destination commune (workplace for active workers, school commune by level for students, residence for the inactive) from the commuting matrices, and finally a precise building within that commune, sampled proportionally to the number of workers (for companies) or floor area (for schools).
\\

\paragraph*{Workplace and school building assignment}

Once each individual has been assigned an activity destination municipality, a precise physical location is attributed to each individual depending on their activity status.

For employed workers, a workplace building is drawn from the SIRET establishment database, which records the precise geographic location and principal activity (NAF code) of every registered French establishment. Within each destination municipality, the probability of being assigned to a given establishment is proportional to a pre-computed weight computed in the preprocessing part. A fraction of employed workers is additionally assigned to school buildings as teaching staff: for each school, the required number of adult workers is estimated from the number of enrolled students and a fixed pupil-to-teacher ratio (18.2 for nursery and primary schools, 13.1 for middle and high schools, 15.5 for universities, estimated from \cite{menesr2025chiffres}  and \cite{bideault2014personnels}).

For students, a school building is drawn from the establishment database stratified by school type, itself determined by age: \textit{maternelle} (ages 3–5), \textit{ecole primaire} (ages 6–10), \textit{collège} (ages 11–14), \textit{lycée} (ages 15–17), and \textit{université} (ages 18 and above). Within each destination municipality and school type, the probability of assignment to a given building is proportional to a weight pre-computed in the preprocessing part. Inactive individuals are assigned their own dwelling as their activity location.
\\

\paragraph*{Work schedule assignment}
Each employed individual is additionally assigned a work schedule type, encoding whether they work nights, Saturdays, Sundays, and/or part-time. For each work category, the probability of working nights, Saturdays, Sundays, or part-time is derived from the French working conditions survey (\textit{Data Source 18}). Each of the four schedule dimensions is then drawn independently as a Bernoulli trial with the corresponding work type level probability, yielding one of 16 possible schedule types encoding all combinations of the four binary dimensions. Students are uniformly assigned schedule type A (standard daytime, weekday-only schedule), while inactive individuals receive no schedule assignment.

\paragraph*{Output encoding.}
The French population is released in the agents' schema common to the three countries (\autoref{tab:output_fields}), with the following country-specific features. The finest residence unit is the municipality: \texttt{cell\_group} equals \texttt{code}, and the exact position of each agent is given by the building of its dwelling. The \texttt{sector} of employed agents is derived from the main activity code (NAF) of the assigned establishment: 2 for NAF divisions 01--43 and 58--75, and 3 for all other divisions, including school staff; it is not assigned to unoccupied agents nor to students. Household types are assigned from the members actually living in each dwelling, and individuals living in collective establishments form one-person households of type 9 (collective housing). Finally, the work schedule type described above is released in an additional, France-specific field, \texttt{timetable}, whose labels encode the combination of night, Saturday, Sunday and part-time work. Students are labelled A, while unoccupied agents have no label.

\subsection*{Italy}
\begin{figure}[ht]
\centering
\includegraphics[width=0.5\textwidth]{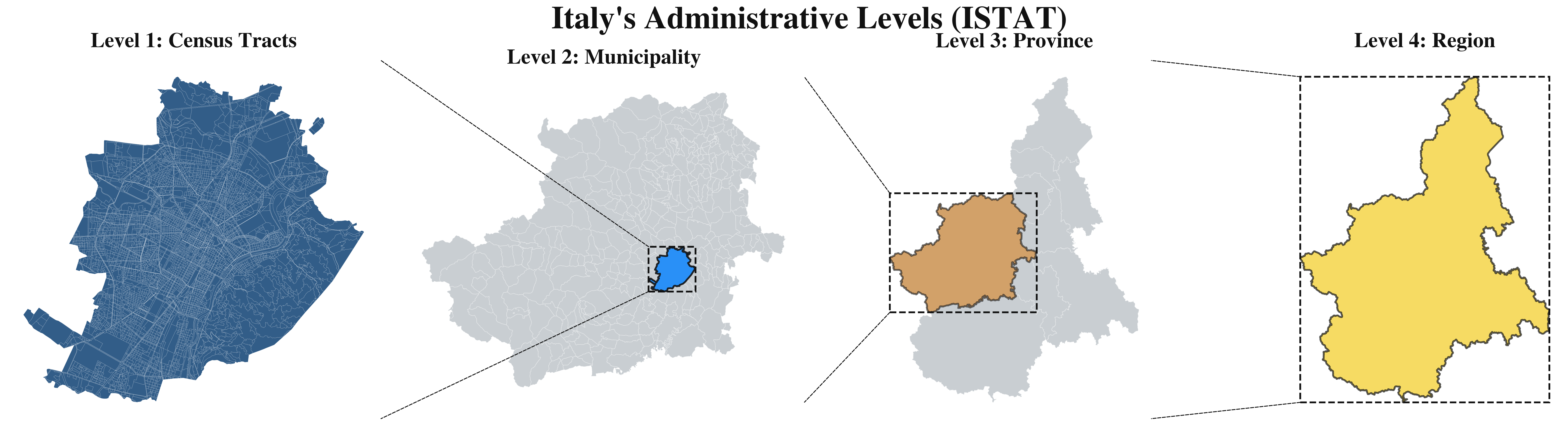}
\caption{Italy's administrative levels, from smallest to largest: Census Tract (sezione di censimento); Municipality (comune); Province (provincia); Region (regione).}
\label{fig:italy_admin}
\end{figure}
In this section, we show the methods to build the agent-based model for Italy. The process can be divided into three parts. In the first part, agents are assigned an age and gender, and then assigned to a home and family. A home features location in the territory and its size. In the second part, agents are assigned an activity and a place to carry the activity out. In the third part, they are assigned a schedule.

\subsubsection*{Definitions}
Italy's territorial organization follows a four-level hierarchy (see Fig.~\ref{fig:italy_admin}). \textbf{Census Tracts} (\textit{sezioni di censimento}) are the smallest statistical units; \textbf{municipalities} (\textit{comuni}) are the fundamental units of local administration;~\textbf{provinces} (\textit{province}) serve as intermediate administrative bodies;~\textbf{regions} (\textit{regioni}) exercise specific autonomy in governance and legislation.
Overall in Italy, we identified $N_t = 451{,}307$ census tracts, $N_m \sim 7.9 \times 10^3$ municipalities, $N_p = 108$ provinces and $N_r = 20$ regions.

\subsubsection*{Households generation}
Each census tract (see~\ref{dat:census_tract} in~\autoref{tab:ita_data_sources}) features a polygon that defines its territorial boundaries, the resident population by age group and sex as well as the number of households by size. We extracted the centroid of each census tract polygon as $\vec{r}_t = (x_t, y_t)$.

We define agent categories based on gender and age intervals. Each agent is assigned a gender (male or female) and an age interval ($[0,5),[5,10),\ldots,[70,75),\geq 75$), yielding a total of $N_a = 2 \times 16 = 32$ distinct agent categories, indexed by $a \in \{1, \ldots, N_a\}$. Dwelling sizes are classified as $1,2,3,4,5,\geq6$ individuals, resulting in $N_s = 6$ possible dwelling size categories, indexed by $s \in \{1, \ldots, N_s\}$.

We extracted the dwelling size matrices $\mat{D} \in \mathbb{N}^{N_t \times N_s}$ and agent matrices $\mat{A} \in \mathbb{N}^{N_t \times N_a}$, where $N_t$ is the number of census tracts. The element $D_{ts}$ represents the number of dwellings of size $s$ in census tract $t$, and $A_{ta}$ represents the number of agents of category $a$ residing in census tract $t$.

A household configuration $f$ denotes the agents sharing the same dwelling, \emph{e.g.} a family living in the same dwelling. It is defined by the vector $\vec{h}_f \in \mathbb{N}^{N_a}$, where $h_{f,a}$ denotes the number of agents of category $a$ in the household. The dwelling size associated to the household configuration is the sum of the elements of $\vec{h}_f$, \emph{i.e.} the total number of agents sharing in the household, with a maximum size of $N_s = 6$. We use the symbol $|\vec{h}_f| = \min(\sum_{a=1}^{N_a} h_{f,a}, N_s)$ to indicate this quantity.

Microdata on household composition (see~\ref{dat:microdata} in~\autoref{tab:ita_data_sources}) provides the distribution of household configurations. We derived $\pi(\vec{h}_f|s)$ as the observed frequency of configuration $\vec{h}_f$ among households of size $s = |\vec{h}_f|$, ensuring that $\sum_{(\vec{h}_f: |\vec{h}_f| = s)} \pi(\vec{h}_f|s)= 1$ for each size $s$. From this data, we identified $N_f \approx 2.5 \times 10^4$ distinct household configurations.

The expected number of households of configuration $\vec{h}_f$ within its own macroregion is $H_f = \pi(\vec{h}_f | |\vec{h}_f|) \sum_{t \,\in\, \mathrm{macroregion}(f)} D_{t,|\vec{h}_f|} \in \mathbb{N}$, where the sum is restricted to census tracts $t$ in the same macroregion as $f$ and gives the total number of dwellings of size $|\vec{h}_f|$ in that macroregion.

For each census tract $t$, we define a census tract vector $\vec{c}_t \in \mathbb{N}^{N_f}$, where $c_{t,f}$ represents the number of households of configuration $\vec{h}_f$ in census tract $t$. We construct auxiliary matrices $\mat{C}_A \in \mathbb{N}^{N_a \times N_f}$ and $\mat{C}_D \in \mathbb{N}^{N_d \times N_f}$, where $C_{A,af} = h_{f,a}$ (the number of agents of type $a$ in configuration $f$) and $C_{D,sf} = 1_{|\vec{h}_f| = s}$ (an indicator for whether configuration $f$ has size $s$). These matrices allow us to aggregate household configurations: $(\mat{C}_A \cdot \vec{c}_t)_a$ gives the total number of agents of type $a$ in census tract $t$, and $(\mat{C}_D \cdot \vec{c}_t)_s$ gives the total number of dwellings of size $s$.

Reconstructing the household-configuration vector $\vec c_t$ for every census tract $t$, so that it is simultaneously consistent with the tract's own counts ($A_{t\cdot}$, $D_{t\cdot}$) and with the macroregional target $\vec H$, is formulated as a constrained-allocation problem and solved tract by tract, rather than as a single national optimization; every tract's target $\vec H$ and candidate configurations are scoped to its own macroregion, as above.

Most census tracts require no optimization at all: a tract whose reported population reduces to a single household, or in which every dwelling is single-occupant, has its configuration fully determined by the tract's own counts and is resolved directly, without a solver call. Of the census tracts that reach this stage, 85.8\% require the full procedure below; the remainder are resolved either trivially by the two criteria above, or --- for tracts whose reported population exceeds their reported dwelling capacity --- under a one-sided relaxation of the dwelling-count constraint below.

For every other tract $t$, with running macroregional counts $p_f$ already placed by tracts processed so far, we solve the integer program
\begin{align}
  \min_{\vec c_t,\,\vec u_t,\,\vec e^{-},\,\vec e^{+}} \quad
  & \sum_{f\in D_t} u_{t,f} \;+\; \lambda\sum_k\bigl(e_k^{-}+e_k^{+}\bigr) \\
  \text{s.t.}\quad
  & (\mat{C}_A \cdot \vec{c}_t)_a + e_{a}^{-} - e_{a}^{+} = A_{ta} && \forall a, \\
  & (\mat{C}_D \cdot \vec{c}_t)_s + e_{s}^{-} - e_{s}^{+} = D_{ts} && \forall s, \\
  & u_{t,f} - c_{t,f} \geq p_f - H_f && \forall f\in D_t, \\
  & u_{t,f} + c_{t,f} \geq H_f - p_f && \forall f\in D_t,
\end{align}
where $c_{t,f}\in\mathbb N$ is the number of households of configuration $f$ allocated to tract $t$, $u_{t,f}\geq 0$ linearizes the distance $\lvert c_{t,f}-(H_f-p_f)\rvert$ to the current macroregional residual, and $e^-,e^+\geq 0$ are elastic slacks on the demographic and dwelling-size constraint rows. At $\lambda=0$ these slacks are pinned to zero and every row is enforced exactly (except the one-sided relaxation on dwelling-count rows noted above); $\lambda$ is switched to a large positive value only at the third tier below.

The candidate domain $D_t$ is built by a three-tier escalation, each tier attempted only if the previous one returns no feasible solution: (1) a residual-weighted sample of configurations that are \emph{structurally admissible} in $t$ (i.e.\ not exceeding any of the tract's own counts in any category), solved with $\lambda=0$, retried with a fresh sample on infeasibility up to a fixed number of attempts; (2) the full macroregional configuration set, unsampled, still with $\lambda=0$; (3) a residual-weighted sample of the \emph{unrestricted} configuration set (no admissibility filter), solved with $\lambda>0$, guaranteed feasible since the elastic slacks can absorb any remaining violation.

\begin{figure}
    \centering
    \begin{subfigure}[b]{0.9\linewidth}
        \centering
        \includegraphics[width=\linewidth]{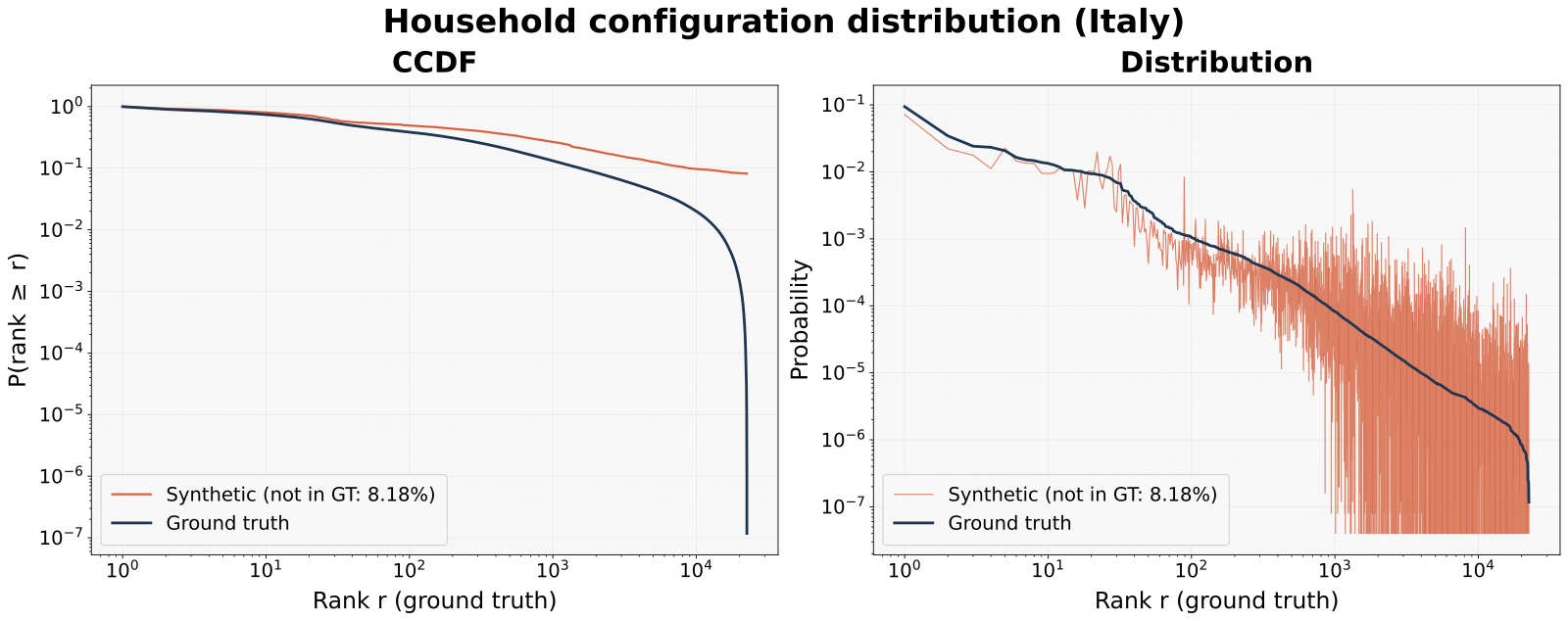}
        \caption{After initialization, no sweeps}
        \label{fig:rank_og_ccdf}
    \end{subfigure}

    \vspace{0.5em}

    \begin{subfigure}[b]{0.9\linewidth}
        \centering
        \includegraphics[width=\linewidth]{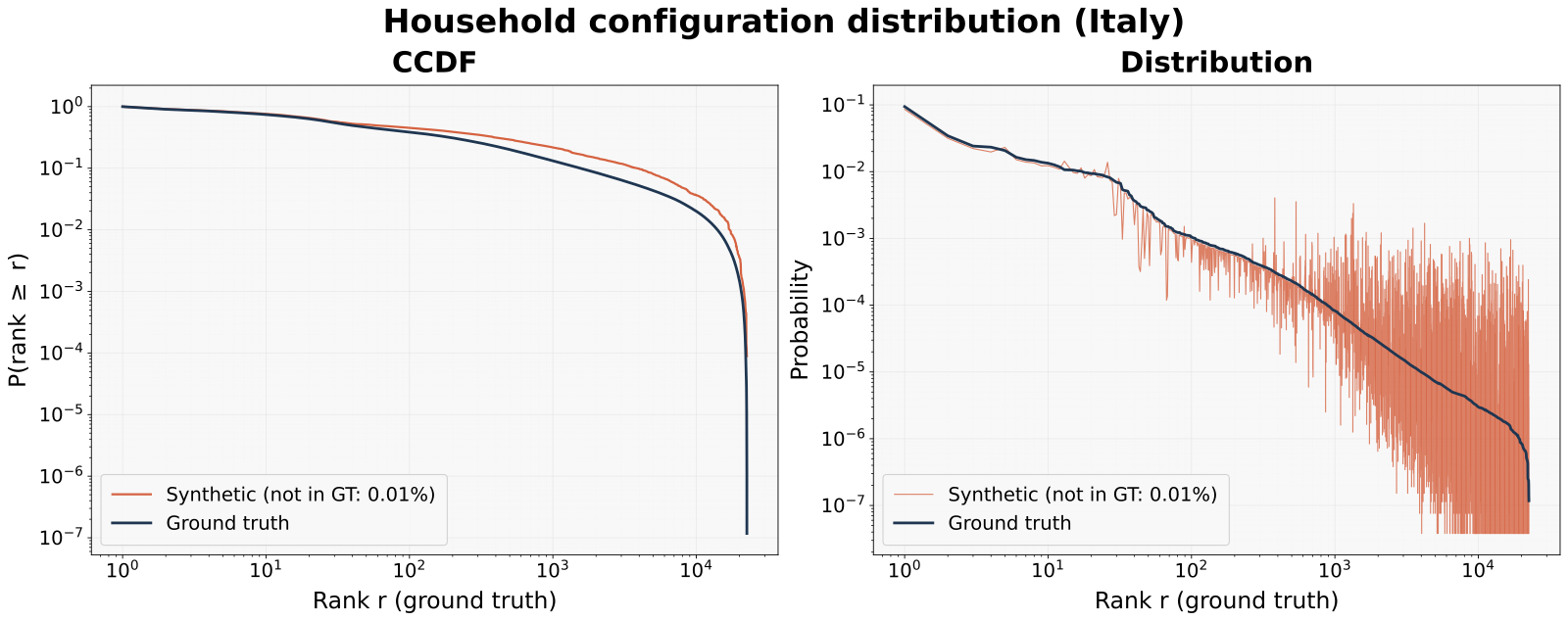}
        \caption{After sweeps}
        \label{fig:rank_new_ccdf}
    \end{subfigure}

    \caption{Comparison of the household-configuration distributions before and after the Metropolis--Hastings sweeps.}
    \label{fig:rank_ccdf_comparison}
\end{figure}

Once every tract has an initial $\vec c_t$ from the pass above, tracts are revisited in repeated sweeps, each in a fresh randomized order, to correct for the order-dependence of that first pass. In each sweep, a tract's current contribution is temporarily removed from $\vec p$, its program is re-solved against the residual this implies, and the move is accepted via a Metropolis--Hastings rule: an improving move (lower objective value) is always accepted, a worsening one is accepted with probability $\exp(-\Delta/T)$ and otherwise rejected. The temperature $T$ is reduced by a fixed factor after every sweep, and the sweep loop stops when the macroregional residual $\lVert \vec H - \vec p \rVert_1$ falls below a tolerance, a full sweep produces no improving move, or a sweep budget is reached. Every solve, at every tier and in both passes, is dispatched to the same open-source mixed-integer solver (HiGHS~\cite{huangfu2018parallelizing}).
As shown in~\autoref{fig:rank_ccdf_comparison}, the rank distribution of household configurations changes noticeably between the initial allocation and the state after 10 Metropolis--Hastings sweeps.

With the obtained census tract vectors $\vec{c}_t$, each individual agent is assigned a gender, age, household, dwelling, and territorial position $\vec{r}_t$.
In particular, the age is extracted from the age range of the agent to respect the age pyramid from that municipality level (see~\ref{dat:pyramid} from~\autoref{tab:ita_data_sources}).

\subsubsection*{Activity assignation}

In this section, we describe how each agent is assigned a daily activity and the corresponding location where the activity is performed.

\paragraph*{Data sources.}
Municipal-level records on professional conditions (see~\ref{dat:professional_condition} in~\autoref{tab:ita_data_sources}) and employment by macrosector (see~\ref{dat:working_stats} in~\autoref{tab:ita_data_sources}) provide distributions of employment status disaggregated by gender, age, and municipality of residence, as well as the distribution of job types by gender and municipality. In this context, ISTAT data provides the distribution on $N_w = 6$ different macrosectors.
The industry census (see~\ref{dat:localunits} in~\autoref{tab:ita_data_sources}) provides the number of local units (workplaces) by economic sector according to ATECO codes, along with the number of employees in each unit. It is possible to map the ATECO code of each local unit into one of the $N_w = 6$ macrosectors.
These data are aggregated at the census tract level, and we assign each local unit a position $\vec{r}_v$ equal to the centroid of its census tract. The census on the schools (see~\ref{dat:schools},\ref{dat:students} in~\autoref{tab:ita_data_sources}) provides the schools location, their educational level and the number of students.
The origin-destination commuting survey (see~\ref{dat:od} in~\autoref{tab:ita_data_sources}) provides origin-destination matrices disaggregated by activity type.

\paragraph*{Professional condition and activity type.}
From these data sources, we derived the matrix $\mat{P} \in [0,1]^{N_a \times N_m \times N_c}$, where $P_{amc}$ represents the probability that an agent of type $a$ residing in municipality $m$ has professional condition $c \in \{\text{employed}, \text{student}, \text{unemployed}\}$. For employed agents, we derived the matrix $\mat{W} \in [0,1]^{N_a \times N_m \times N_w}$, where $W_{amw}$ gives the probability that an employed agent of type $a$ in municipality $m$ works in macrosector $w \in \{1,\ldots,N_w\}$. 

Students are assigned to educational levels based solely on age: agents aged $[0, 6)$ attend preschool, $[6, 11)$ primary school, $[11, 14)$ lower secondary school, $[14, 19)$ upper secondary school, and $\geq 19$ university. To the best of our knowledge, detailed data on educational enrollment by age are not available at the required spatial resolution; therefore, we use this deterministic age-based assignment.

\paragraph*{Spatial allocation.}\label{par:spatial}
From the origin-destination survey, we derived the matrix $\mat{O} \in [0,1]^{N_m \times N_m \times N_v}$, where $O_{m_d m_o v}$ represents the probability of traveling from origin municipality $m_o$ to destination municipality $m_d$ for activity $v$. The activity type $v$ depends on professional condition: for students, $v$ corresponds to an educational level; for employed agents, $v$ corresponds to a work macrosector $w$.

For unemployed agents, we set $O_{m_d m_o \text{unemployed}} = \delta(m_d, m_o)$, meaning they remain in their municipality of residence. Since the commuting survey was administered only to upper secondary and university students, we similarly set $O_{m_d m_o e} = \delta(m_d, m_o)$ for $e \in \{\text{preschool}, \text{primary}, \text{lower secondary}\}$, assuming younger students attend schools within their municipality of residence.

We ensured that the matrix $\mat{O}$ satisfies the following constraints: (i) $\sum_{m_d} O_{m_d m_o v} = 1$ for all origin municipalities $m_o$ and activities $v$; and (ii) if $O_{m_d m_o v} > 0$, then at least one unit (school or workplace) exists in destination municipality $m_d$ where activity $v$ can be performed. In cases where no suitable destination exists (i.e., $\sum_{m_d} O_{m_d m_o v} = 0$), we iteratively search neighboring municipalities until finding at least one municipality with the required activity type, then assign equal probability to all such destinations.

\paragraph*{Assignment procedure.}
To assign an activity and destination to each agent, we proceed as follows: (1) sample a professional condition $c$ from $\mat{P}$ based on the agent's type $a$ and municipality of residence $m$; (2) if $c = \text{employed}$, sample a work macrosector from $\mat{W}$; if $c = \text{student}$, assign an educational level based on age; (3) sample a destination municipality $m_d$ from $\mat{O}$; and (4) assign a specific workplace or school within $m_d$, with selection probability proportional to the size (number of employees or students) of each unit of the appropriate type.

Following this procedure, each agent features demographic attributes (gender, age, household, residence) from the previous step, and an activity type with associated location (workplace or school) from this step.

\subsubsection*{Schedules}

Official statistics on working hours disaggregated by macrosector are not available at the national level. To address this gap, we combined information from two sources: (i) operating hours extracted from OpenStreetMap (OSM) building tags, which provide opening and closing times for commercial and service establishments, and (ii) typical work schedules documented in collective bargaining agreements published by major trade unions. When multiple shift-based schedules are available for a given macrosector, we sample uniformly among them, as no finer-grained data are available to inform their relative frequencies.

 




%% file: data_source.tex
\subsection*{Common Data sources}

\textbf{OpenStreetMap (OSM)} We used OpenStreetMap data to complement the available public datasets of each country. OSM provides a granular classification of building types across countries, whereas public data from public authorities, e.g. cadastral data in Spain, broadly categorizes buildings into six functional types (agricultural, industrial, office, retail, public services, and residential). OSM includes a rich and detailed taxonomy, such as:
\noindent \textbf{Public facilities:} government buildings, police stations; \\
\noindent\textbf{Healthcare services:} hospitals, clinics, pharmacies; \\
\noindent\textbf{Commercial buildings:} restaurants, supermarkets, banks, retail outlets; \\
\noindent\textbf{Recreational and special-purpose structures}.
This detailed classification is critical for assigning daily activities to individuals in the synthetic population, enabling realistic simulation of commuting, education, shopping, healthcare usage, and leisure behaviors.

The OSM data were spatially integrated with the cadastral dataset to produce a harmonized building inventory enriched with attributes such as primary function, floor area, geolocation, and usage category.

\subsection*{Data sources for Spain}

The synthetic population for Spain was constructed using three primary data sources: cadastral data, OpenStreetMap (OSM), and demographic and mobility statistics from the Spanish National Statistics Institute (INE). These datasets collectively enable the calibration of key features of the synthetic population, including household composition, age structure, and mobility behavior.

\textbf{Cadastral Data (INSPIRE)} The first data source is the INSPIRE-compliant cadastral dataset \cite{catastro_inspire_services}, retrieved from the Spanish Cadastre through its ATOM feed service. This dataset serves two essential functions: enabling the spatial assignment of individuals to specific residential buildings and identify structures associated with key activities, including work, education, and healthcare.
The dataset includes information on the constructed area of each building, which is used to estimate occupancy levels and contributes to downstream modeling processes such as the estimation of transmission risk in indoor environments.
Although the INSPIRE schema defines three layers (Addresses, Parcels, and Buildings), only the \textbf{Buildings} layer was used in this study, as it provides the structural and spatial information required. Building data were systematically collected for all Spanish municipalities, with the exception of \textit{Navarra} and the \textit{Basque Country}, which maintain independent cadastral registries. Data for these regions will be incorporated separately from their respective sources.

\textbf{INE Demographic and Mobility Statistics} The third major data source is the \textit{Instituto Nacional de Estadística} (INE), which provides demographic, household, labor force, and mobility datasets. \\
\noindent \textbf{Population and Housing Census Data (2021):} Includes municipal-level population counts and dwelling data. For municipalities with populations over 2.000, it also reports the number of unoccupied dwellings.
\noindent \textbf{Household microdata:} surveys providing detailed anonymized information on household composition, age, gender, education, employment, income, and housing conditions.  \\
\noindent \textbf{Labor Force microdata} (\textit{Encuesta de Población Activa}, EPA): surveys including individual-level labor force indicators such as employment status, occupation, education, and hours worked.
These microdata inform the synthetic population's demographic and socioeconomic profiles. The specific attributes retained after preprocessing are listed in \textit{Annex A}. \\
\noindent \textbf{Mobility Data:} origin-destination aggregated flows at nationwide resolution of a grid of spatial cells covering all of Spain.  
These mobility data are used to simulate realistic daily travel and spatial interactions between individuals in the synthetic population.

\subsection*{Data Sources for France}

The synthetic French population is constructed using three main sources of data: demographic surveys from the National Institute of Statistics and Economic Studies (INSEE), microdata provided by Progedo (a French public data infrastructure), and the national building database (BDND).

\textbf{National Institute of Statistics and Economic Studies (INSEE)} Based on census data, INSEE produces sub-municipality (IRIS level) and municipality level distributions: the number of households by type, the age, sex, socio-professional categories of the reference person according to household type, the number of children by age of the reference person, income of the reference person, and the main activity (employed, student, inactive) by age. INSEE also provides data on the number of individuals living in a given municipality and working or studying in another. Finally, we also use age-sex pyramid at IRIS level (numbre of people by sex and age catgory in each IRIS).

\noindent \textbf{National Building Database (BDND)} The National Building Database provides a comprehensive dataset of all buildings in France. For each building, the database includes: its geometry (as a multipolygon), surface area, geographic location (longitude and latitude), number of floors, number of housing units, and nearby points of interest (such as schools, shops, medical practices, hospitals, post offices, nursing home, student housing etc.).

\noindent \textbf{SIRET Database} The SIRET database contains all enterprises buildings in France, along with their geographic location and the number of employees they host (reported in size brackets).

\noindent \textbf{Microdata from Progedo}
We used a microdata sample (i.e., an anonymized subset of the household population) that includes the age of the reference person and the age of their partner.

\subsection*{Data Sources for Italy}

The synthetic Italian population is constructed using demographic and socioeconomic data from the Italian National Institute of Statistics (ISTAT) and educational data from the Italian Ministry of Public Education (MIM).

\textbf{Italian National Institute of Statistics (ISTAT)} provides comprehensive demographic, household, and employment data at multiple spatial resolutions. The key datasets include:

\noindent\textbf{Household composition microdata (2001):} A sample of approximately $2\%$ of the population used to reconstruct the distribution of household configurations by demographic composition.

\noindent \textbf{Census tract records (2021):} Complete territorial coverage of residents disaggregated by age and gender, and households by size.

\noindent \textbf{Industry census (2011):} Data on local units (workplaces) including their location, economic sector, and number of employees.

\noindent \textbf{Origin-destination commuting survey (2011):} Matrices describing commuting flows between municipalities for work and education purposes.

\noindent \textbf{Administrative boundaries:} Shapefiles for census tracts, municipalities, provinces, regions, and macroregions, along with the current list of municipalities (2025) and historical records of territorial variations (1991--2025).

\textbf{The Italian Ministry of Public Education (MIM)} provides detailed information on educational facilities:

\noindent \textbf{School registry:} Complete list of schools with geocoded addresses and education level (preschool, primary, lower secondary, upper secondary).

\noindent \textbf{School enrollment:} Number of students enrolled in each school.

A comprehensive description of all datasets, including access URLs and processing details, is provided in~\autoref{tab:ita_data_sources}.
